\documentclass[reprint, aps, prd, letterpaper, noshowpacs, amsmath, %
amssymb, amsfonts, nofootinbib, floatfix, superscriptaddress, %
twoside]{revtex4-1}
\pdfoutput=1

\usepackage{enumerate}

\usepackage[T3,T1]{fontenc}
\usepackage{mathrsfs} 
\usepackage{bm} 
\makeatletter
\@ifclassloaded{beamer}
  { 
    \typeout{UsePackages: Detected beamer}
    \usepackage{tgheros}
    
  }
  { 
    \typeout{UsePackages: Did not detect beamer}
    \ifx\asybeamer\undefined 
    \typeout{UsePackages: Detected article}
    \usepackage[varg]{txfonts} 
    \usepackage{tgtermes} 
    \else 
    \typeout{Fonts: Detected asy for beamer}
    \usepackage{tgheros}
    
    \fi
  }
\usepackage{microtype} 
\def\MT@register@subst@font{
  \MT@exp@one@n\MT@in@clist\font@name\MT@font@list
  \ifMT@inlist@\else\xdef\MT@font@list{\MT@font@list\font@name,}\fi}
\makeatother

\usepackage{graphicx} %
\usepackage[x11names,svgnames,rgb]{xcolor} %
\usepackage{xspace}
\usepackage{braket}
\usepackage{accents}
\usepackage{siunitx}

\usepackage{mathtools}
\DeclareSymbolFontAlphabet{\mathrm}{operators}

\definecolor{CiteColor}{rgb}{0.18039, 0.18824, 0.57255}
\definecolor{UrlColor} {rgb}{0.741, 0.173, 0.000}
\definecolor{DarkUrlColor} {rgb}{0.500, 0.110, 0.000}
\definecolor{LinkColor}{rgb}{0.25098, 0.47843, 0.04706}

\makeatletter %
\newcommand{\ShowFont}{%
  \typeout{The main font is \f@encoding \space \f@family \space %
    \f@series \space \f@shape \space at \f@size pt.}%
  \typeout{The math font sizes are \tf@size pt (main), \sf@size pt %
    (script), and \ssf@size pt (scriptscript).}%
  \typeout{The linewidth is \the\linewidth}} %
\makeatother %

\usepackage{xargs}

\makeatletter

\def\@seccntformat#1{\csname the#1\endcsname.~}%

\def\section{%
  \@startsection {section}
  {1} {\z@} {0.55cm \@plus1ex \@minus .02ex}%
    {0.225cm} { \normalfont\bfseries \centering}%
}%
\def\subsection{%
  \@startsection {subsection}
  {2} {\z@ } {0.45cm \@plus 0.8ex \@minus 0.2ex}%
  {0.1125cm}{\normalfont \bfseries \centering }}
\def\subsubsection{%
  \@startsection {subsubsection}
  {3} {\z@ } {0.4cm \@plus 0.6ex \@minus 0.1ex}%
  {0.075cm}{\normalfont \it \centering }}

\newcommand{\surnames}[1]{\def\@surnamelist{#1}\relax}
\surnames{\ }
\ifx \@shorttitle \@empty
\else
  \def\@oddhead{\small \MakeUppercase{\@shorttitle} \hfill}
\fi
\ifx \@surnamelist \@empty
\else
  \def\@evenhead{\small \MakeUppercase{\@surnamelist} \hfill}
\fi
\def\@oddfoot{\reset@font\hfil\thepage\hfil}%
\def\@evenfoot{\reset@font\hfil\thepage\hfil}%
\g@addto@macro\maketitle{\global\@specialpagetrue\gdef\@specialstyle{plain}}

\usepackage[colorlinks, plainpages=false,
hyperfigures=true]{hyperref}
\hypersetup{linkcolor=LinkColor}
\hypersetup{citecolor=CiteColor}
\hypersetup{urlcolor=UrlColor}
\hypersetup{setpagesize=false}
\hypersetup{pdfborder=0 0 0}

\makeatother

\let\Originalddefinition\d
\renewcommand{\d}{\ensuremath{\mathrm{d}}}

\newcommand{\e}{\ensuremath{\mathrm{e}}}
\let\Originalidefinition\i
\renewcommand{\i}{\ensuremath{\mathrm{i}}}
\DeclareSymbolFont{wasy}{U}{wasy}{m}{n}
\DeclareMathSymbol{\thorn}{\mathord}{wasy}{105}
\DeclareMathSymbol{\Thorn}{\mathord}{wasy}{106}
\newcommand\Eth{\text{\DH}}
\newcommand\Alpha{\mathrm{A}}
\newcommand\Beta{\mathrm{B}}
\newcommand\Epsilon{\mathrm{E}}
\newcommand\Varepsilon{\mathit{E}}
\newcommand\Zeta{\mathrm{Z}}
\newcommand\Eta{\mathrm{H}}
\newcommand\Vartheta{\varTheta}
\newcommand\Iota{\mathrm{I}}
\newcommand\Kappa{\mathrm{K}}
\newcommand\Mu{\mathrm{M}}
\newcommand\Nu{\mathrm{N}}
\chardef\omicron=111 
\chardef\Omicron=79  
\newcommand{\Varpi}{\varPi}
\newcommand\Rho{\mathrm{P}}
\newcommand\Varrho{\mathit{P}}
\newcommand\Varsigma{\varSigma}
\newcommand\Varphi{\varPhi}
\newcommand\Tau{\mathrm{T}}
\newcommand\Chi{\mathrm{X}}
\newcommand\MakeAllUppercase[1]{%
  \begingroup
  \let\eth\Eth
  \let\thorn\Thorn
  \let\alpha\Alpha
  \let\beta\Beta
  \let\gamma\Gamma
  \let\delta\Delta
  \let\epsilon\Epsilon
  \let\varepsilon\Varepsilon
  \let\zeta\Zeta
  \let\eta\Eta
  \let\theta\Theta
  \let\vartheta\Vartheta
  \let\iota\Iota
  \let\kappa\Kappa
  \let\lambda\Lambda
  \let\mu\Mu
  \let\nu\Nu
  \let\xi\Xi
  \let\omicron\Omicron
  \let\pi\Pi
  \let\varpi\Varpi
  \let\rho\Rho
  \let\varrho\Varrho
  \let\sigma\Sigma
  \let\varsigma\Varsigma
  \let\tau\Tau
  \let\upsilon\Upsilon
  \let\phi\Phi
  \let\varphi\Varphi
  \let\chi\Chi
  \let\psi\Psi
  \let\omega\Omega
  \MakeUppercase{#1}%
  \endgroup
}

\let\Originalcdefinition\c
\renewcommand{\c}{\mathrm{c}}

\newcommand{\MSun}{\ensuremath{M_\odot}\xspace}

\DeclareSIUnit{\strain}{strain}
\DeclareSIPrePower{\root}{1/2}
\DeclareSIUnit{\parsec}{pc}
\DeclareSIUnit{\yr}{yr}
\DeclareSIUnit{\year}{yr}
\DeclareSIUnit{\lightyear}{ly}
\DeclareSIUnit{\SolarMass}{\ensuremath{\MSun}}
\DeclareSIUnit{\Mass}{\ensuremath{M}}

\newcommand{\defined}{\coloneqq}

\newcommand{\isomorphic}{\approx}

\newcommand{\scriplus}{\ensuremath{\mathscr{I}^{+}}}

\DeclareSymbolFont{tipa}{T3}{tipa}{m}{n}
\DeclareMathAccent{\ibreve}{\mathalpha}{tipa}{'020}

\newcommandx{\structure}[3][2={}, 3={}]{f^{#1 #2}_{#3}}

\newcommand{\SU}[1]{\ensuremath{\mathrm{SU}(#1)}}

\newcommand{\Spin}[1]{\ensuremath{\mathrm{Spin}(#1)}}

\makeatletter
\newcommand{\foreign}[1]{\textit{#1}} 
\newcommand{\etal}{\foreign{et~al}\@ifnextchar{\relax}{.\relax}{\ifx\@let@token.\else\ifx\@let@token~.\else.\@\xspace\fi\fi}}
\newcommand{\etc}{\foreign{etc}\@ifnextchar{\relax}{.\relax}{\ifx\@let@token.\else\ifx\@let@token~.\else.\@\xspace\fi\fi}}
\newcommand{\eg}{\foreign{e.g}\@ifnextchar{\relax}{.\relax}{\ifx\@let@token.\else\ifx\@let@token~.\else.\@\xspace\fi\fi}}
\newcommand{\ie}{\foreign{i.e}\@ifnextchar{\relax}{.\relax}{\ifx\@let@token.\else\ifx\@let@token~.\else.\@\xspace\fi\fi}}

\newcommand{\cf}{\foreign{cf}\@ifnextchar{\relax}{.\relax}{\ifx\@let@token.\else\ifx\@let@token~.\else.\@\xspace\fi\fi}}
\makeatother

\definecolor{NoteColor}{rgb}{0.900, 0.218, 0.000}

\definecolor{NoteQuietColor}{rgb}{0.300, 0.073, 0.000}

\definecolor{NewColor}{rgb}{0,.55,0}

\makeatletter

\newcommand{\ShowDimensions}{%
  \typeout{The font encoding is \f@encoding}                  %
  \typeout{The font family is \f@family}                      %
  \typeout{The font series is \f@series}                      %
  \typeout{The font shape is \f@shape}                        %
  \typeout{The font size is \f@size}                          %
  \typeout{The baselineskip is \f@baselineskip}               %
  \typeout{The math font size is \tf@size}                    %
  \typeout{The math script size is \sf@size}                  %
  \typeout{The math scriptscript size is \ssf@size}           %
  \typeout{The linewidth is \the\linewidth}                   %
  \typeout{The textwidth is \the\textwidth}                   %
  \typeout{The slant is \the\fontdimen1\font}                 %
  \typeout{The inter word space is \the\fontdimen2\font}      %
  \typeout{The inter word stretch is \the\fontdimen3\font}    %
  \typeout{The inter word shrink is \the\fontdimen4\font}     %
  \typeout{The extra space is \the\fontdimen7\font}           %
  \typeout{The xspace skip is \the\xspaceskip}                %
  \typeout{The hyphenation character is \the\hyphenchar\font} %
}

\newcommand{\prefixscripts}[2]{%
  \@mathmeasure\z@\displaystyle{#2}%
  \global\setbox\@ne\vbox to\ht\z@{}\dp\@ne\dp\z@
  \setbox\tw@\box\@ne
  \@mathmeasure4\displaystyle{\copy\tw@#1}%
  \@mathmeasure6\displaystyle{#2}%
  \dimen@-\wd6 \advance\dimen@\wd4 \advance\dimen@\wd\z@
  \hbox to\dimen@{}{\kern-\dimen@\box4\box6}%
}

\newcommand{\scripts}[3]{%
  \@mathmeasure\z@\displaystyle{#2}%
  \global\setbox\@ne\vbox to\ht\z@{}\dp\@ne\dp\z@
  \setbox\tw@\box\@ne
  \@mathmeasure4\displaystyle{\copy\tw@#1}%
  \@mathmeasure6\displaystyle{#2#3}%
  \dimen@-\wd6 \advance\dimen@\wd4 \advance\dimen@\wd\z@
  \hbox to\dimen@{}{\kern-\dimen@\box4\box6}%
}
\makeatother

\RequirePackage{braket}
\makeatletter
\DeclareFontFamily{OMX}{MnSymbolE}{}
\DeclareSymbolFont{MnLargeSymbols}{OMX}{MnSymbolE}{m}{n}
\SetSymbolFont{MnLargeSymbols}{bold}{OMX}{MnSymbolE}{b}{n}
\DeclareFontShape{OMX}{MnSymbolE}{m}{n}{
    <-6>  MnSymbolE5
   <6-7>  MnSymbolE6
   <7-8>  MnSymbolE7
   <8-9>  MnSymbolE8
   <9-10> MnSymbolE9
  <10-12> MnSymbolE10
  <12->   MnSymbolE12
}{}
\DeclareFontShape{OMX}{MnSymbolE}{b}{n}{
    <-6>  MnSymbolE-Bold5
   <6-7>  MnSymbolE-Bold6
   <7-8>  MnSymbolE-Bold7
   <8-9>  MnSymbolE-Bold8
   <9-10> MnSymbolE-Bold9
  <10-12> MnSymbolE-Bold10
  <12->   MnSymbolE-Bold12
}{}

\let\llangle\@undefined
\let\rrangle\@undefined
\DeclareMathDelimiter{\llangle}{\mathopen}%
                     {MnLargeSymbols}{'164}{MnLargeSymbols}{'164}
\DeclareMathDelimiter{\rrangle}{\mathclose}%
                     {MnLargeSymbols}{'171}{MnLargeSymbols}{'171}
\let\protect\relax

{\catcode`\|=\active
  \xdef\InnerProduct{\protect\expandafter\noexpand\csname InnerProduct \endcsname}
  \expandafter\gdef\csname InnerProduct \endcsname#1{%
    \begingroup
    \ifx\SavedDoubleVert\relax
    \let\SavedDoubleVert\|\let\|\IpDoubleVert
    \fi
    \mathcode`\|32768\let|\IPVert
    \left({#1}\right)
    \endgroup
  }
  \xdef\Bbrakket{\protect\expandafter\noexpand\csname Bbrakket \endcsname}
  \expandafter\gdef\csname Bbrakket \endcsname#1{\begingroup
     \ifx\SavedDoubleVert\relax
       \let\SavedDoubleVert\|\let\|\BbraDoubleVert
     \fi
     \mathcode`\|32768\let|\BraVert
     \left\llangle{#1}\right\rrangle\endgroup}
}
\def\IPVert{\@ifnextchar|{\|\@gobble}
     {\egroup\,\mid@vertical\,\bgroup}}
\def\IPDoubleVert{\egroup\,\mid@dblvertical\,\bgroup}
\let\SavedDoubleVert\relax
\def\midvert{\egroup\mid\bgroup}
\def\SetVert{\@ifnextchar|{\|\@gobble}
    {\egroup\;\mid@vertical\;\bgroup}}
\def\SetDoubleVert{\egroup\;\mid@dblvertical\;\bgroup}
\def\mid@vertical{\mskip1mu\vrule\mskip1mu}
\def\mid@dblvertical{\mskip1mu\vrule\mskip2.5mu\vrule\mskip1mu}
\makeatother

\newcommand{\Cornell}{\affiliation{Cornell Center for Astrophysics and
    Planetary Science, Cornell University, Ithaca, New York 14853,
    USA}} %

\newcommand{\CITA}{\affiliation{Canadian Institute for Theoretical
    Astrophysics, University of Toronto, 60 Saint George Street,
    Toronto, Ontario M5S 3H8, Canada}} %

\newcommand{\AEI}{\affiliation{Albert-Einstein-Institut,
    Max-Planck-Institut f{\"{u}}r Gravitationsphysik, D-14476
    Potsdam-Golm, Germany}} %

\usepackage{xspace}
\usepackage{color}
\usepackage[normalem]{ulem} 
\usepackage{color,soul}

\begin{document}


\graphicspath{%
  {Plots/}%
}

\title[Compact binary waveform c.m. corrections] {Compact binary
  waveform center-of-mass corrections}

\makeatletter \@booleantrue\frontmatterverbose@sw \makeatother

\author{Charles J. Woodford} \CITA
\author{Michael Boyle} \Cornell
\author{Harald P. Pfeiffer} \CITA \AEI

\date{\today}

\begin{abstract}
  We present a detailed study of the center-of-mass (c.m.) motion seen
  in simulations produced by the Simulating eXtreme Spacetimes (SXS)
  collaboration. We investigate potential physical sources for the
  large c.m. motion in binary black hole simulations and find
  that a significant fraction of the c.m. motion cannot be explained
  physically, thus concluding that it is largely a gauge effect. These
  large c.m. displacements cause mode mixing in the gravitational
  waveform, most easily recognized as amplitude oscillations caused by
  the dominant $(2, \pm 2)$ modes mixing into subdominant modes.  This
  mixing \emph{does not} diminish with increasing distance from the
  source; it is present even in \emph{asymptotic} waveforms,
  regardless of the method of data extraction.  We describe the
  current c.m.-correction method used by the SXS collaboration, which
  is based on counteracting the motion of the c.m. as measured by the
  trajectories of the apparent horizons in the simulations, and
  investigate potential methods to improve that correction to the
  waveform. We also present a complementary method for computing an
  optimal c.m. correction or evaluating any other c.m. transformation
  based solely on the asymptotic waveform data.
\end{abstract}



\maketitle



\section{Introduction}
\label{sec:Introduction}
Binary black hole (BBH) systems have been studied for decades,
beginning with analytic work and branching out into numerical
relativity.  With the introduction of gravitational-wave detectors,
particularly LIGO, the pursuit of BBH gravitational waveforms has
intensified in an attempt to create and fill vast waveform template
banks. Gravitational waveforms created through numerical relativity
are generally the most accurate waveforms available, and are used for
parameter estimation and to compare and improve semianalytic and
analytic models of BBHs, which in turn are used for gravitational-wave
detection and parameter estimation~\cite{Lovelace2016, Healy2017,
  Kumar2013}.

While numerical-relativity waveforms are the most accurate BBH
waveforms, there are concerns regarding their validity, accuracy, and
reproducibility. There have been numerous discussions on how to
measure the accuracy of a numerical relativity simulation and some
sources of error in the simulations have been investigated, including
numerical truncation errors, error due to extraction at finite radius
or imperfect extrapolation to infinite radius, and errors between
simulations of different lengths that otherwise have identical
parameters~\cite{Blackman2017, Chu2015, PhysRevD.80.124045}.

Though not strictly a source of \emph{error} like those named above,
there are also consequences due to the gauge freedom of general
relativity that may be confused for errors if not properly understood,
and will effectively become sources of error if
ignored~\cite{PhysRevD.78.044024, PhysRevD.87.084004, Boyle2016}.
Gauge freedom in general relativity affects all numerical relativity
simulations, and thus far no numerical relativity results in the
literature have been in a completely specified gauge due to this
inherent gauge freedom.  While a full accounting of the effects of
general gauge freedom is beyond the scope of this work, we will
address the translation and boost degrees of freedom.
Reference~\cite{Boyle2016} identified these transformations as
important for counteracting the observed motion of the center of mass
in simulations produced by the Simulating eXtreme Spacetimes (SXS)
collaboration.  Here, we expand on that analysis, using the recently
updated catalog of 2,018 SXS simulations~\cite{SXSCatalog,
  SXSCatalog2019}, and investigating possible improvements to the
correction method.

A translation $\vec{\alpha}$ and a boost $\vec{\beta}$ will transform
the waveform $h$, measured at some point distant from the source, as
\begin{subequations}
  \label{eq:transformation}
  \begin{align}
    \label{eq:first-order-transformation}
    h(t) \to {}
    &
      h\left(t + (\vec{\alpha} + \vec{\beta}t)\cdot\hat{n} \right)
      + \mathscr{O} \left( \lvert \vec{\beta} \rvert h \right)
      + \mathscr{O} \left( \lvert \vec{\alpha} + \vec{\beta}t
      \rvert^{2} \partial^{2}_{t}h \right), \\
    \label{eq:taylor-expanded-transformation}
    & \approx
      h(t) + \partial_{t}{h}(t)\,
      (\vec{\alpha} + \vec{\beta} t) \cdot \hat{n},
  \end{align}
\end{subequations}
where $\hat{n}$ is the direction to the observer from the
source~\cite{Boyle2016}.  Note that this is independent of the
distance to the source; even the asymptotic waveform will exhibit this
dependence regardless of any extrapolation, Cauchy-characteristic
extraction, or similar techniques that may be applied to the data.  We
can understand this intuitively by thinking about a sphere surrounding
the source.  If we displace the source away from the center of the
sphere, an emitted signal will arrive at the part of the sphere
closest to the source before it will arrive at the opposite side of
the sphere.  The difference in arrival times is independent of the
radius of the sphere; it only depends on the size of the displacement.
The additional term in Eq.~\eqref{eq:taylor-expanded-transformation}
introduces an angular dependence that is not generally included in
waveform models.

\begin{figure*}
  \includegraphics[width=\linewidth]{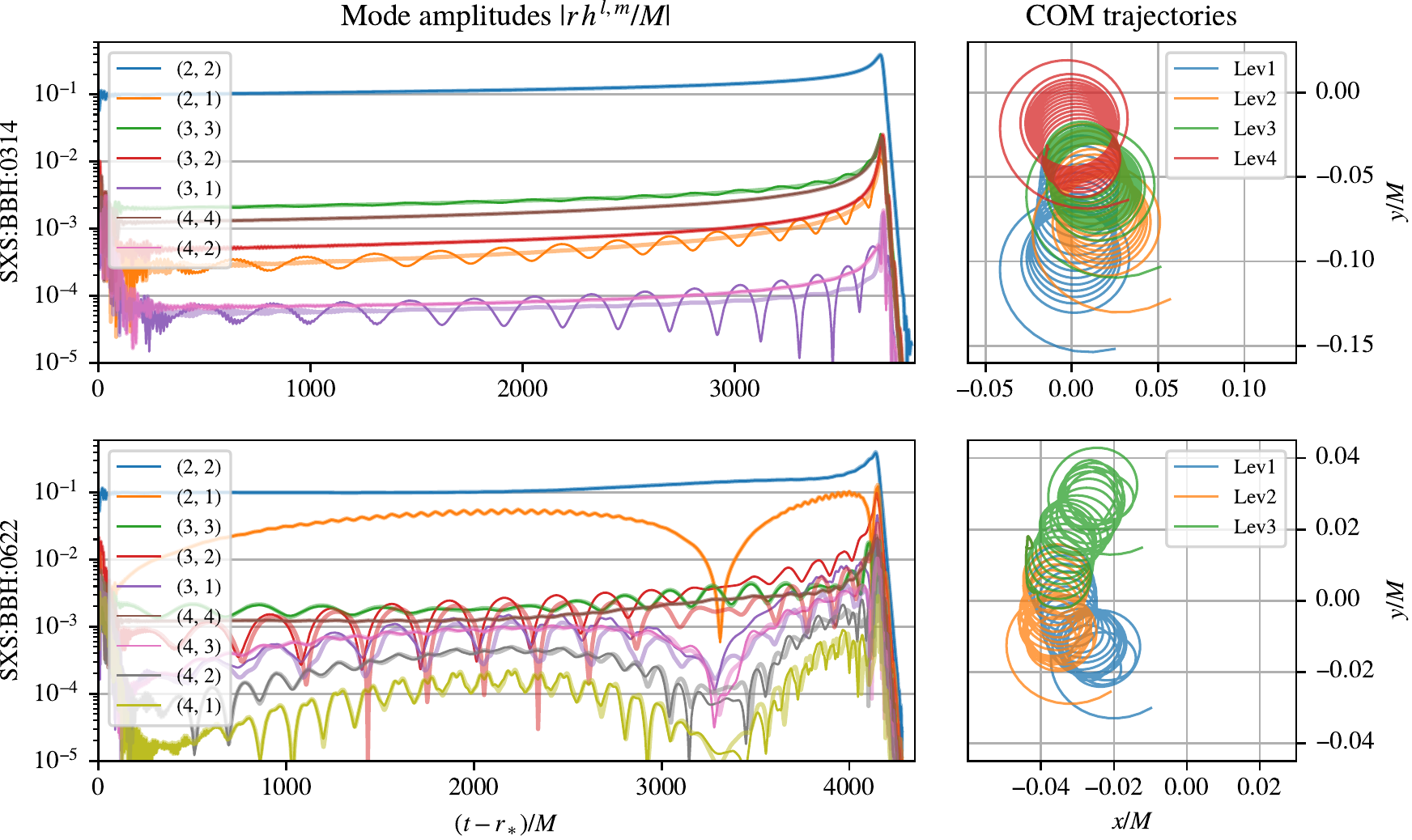}
  \caption{\label{fig:ampAndpos}%
    Center-of-mass motion and its effect on waveforms.  These plots
    show data for two systems from the SXS catalog.  The upper panels
    correspond to the nonprecessing system
    SXS:BBH:0314~\cite{SXS:BBH:0314}, which has a mass ratio of
    $1.23$, with spins of $0.31$ for the larger BH and $-0.46$ for the
    smaller BH, and are aligned with the orbital angular momentum.
    The lower panels correspond to the precessing system
    SXS:BBH:0622~\cite{SXS:BBH:0622}, which has a mass ratio of $1.2$,
    with randomly oriented spins of magnitude $0.85$.  The panels on
    the right side show the c.m. trajectories in the simulation
    coordinates calculated from the apparent horizons of each black
    hole for a variety of resolutions. In addition to the roughly
    circular motion, offsets and drifts are also apparent in each case
    and are the portions of the motion that we remove using inertial
    transformations.  Note that no convergence is evident, reinforcing
    the idea that the motion is effectively random.  The panels on the
    left show the dominant mode amplitudes of each system, both before
    and after c.m. correction---the thin darker lines being the raw
    waveform data, and the thicker transparent lines being the
    corrected data.  In the nonprecessing case (upper panel), we see
    modulations in the raw data that are not expected on physical
    grounds; even the relatively small c.m. motion gives rise to
    clearly visible effects.  No such modulations are visible in the
    corrected data.  In the precessing case (lower panel), modulations
    are present in both the raw and corrected data, caused by mode
    mixing due to the precession of the system itself.  It is not
    obvious from this plot alone that the correction makes any
    improvement to the data.  In Sec.~\ref{sec:Accuracy}, we define a
    quantitative measure of the waveform that very clearly
    distinguishes the corrected data as a significant improvement.}
\end{figure*}

Figure~\ref{fig:ampAndpos} demonstrates the effects of c.m. motion for
two systems from the SXS catalog.  The most striking example is the
upper pair of panels, which show data from a nonprecessing system
with mass ratio 1.23.  On physical grounds, there is nothing to
suggest modulations in the mode amplitudes on the orbital timescale;
this is a relatively symmetric system with very low eccentricity.  The
dominant physical behavior on the orbital timescale is simply
rotation, which should have no effect on the amplitudes of the modes,
along with a secular increase toward merger.  Nonetheless, the raw
waveform data from the simulation (thin dark lines in the upper left
panel) shows very clear amplitude modulations of the subdominant
modes on the orbital timescale.  These modulations---like the c.m.
trajectories seen in the upper right panel---show no signs of
convergence with increasing numerical resolution in the simulation,
even though the initial data for each resolution is created from
identical high-resolution initial data.

As we discuss below, the c.m. motions found in the SXS catalog are
effectively random and apparently independent of any physical
parameters of the systems.  Therefore, they comprise an essentially
random source of unmodeled and unphysical contributions to waveforms
from numerical relativity.  In particular, they are not systematic;
the modulations found in waveforms for one set of physical parameters
will be uncorrelated with the modulations in waveforms even for nearly
identical physical parameters. Clearly, expecting waveform models such
as effective-one-body (EOB) ~\cite{Buonanno1998, Pan2013,
  Taracchini2013, Babak2016, Bohe2017, Cotesta2018},
phenomenological~\cite{Ajith2007, Khan2018, Nagar2018}, and surrogate
models~\cite{Varma2018} to accurately represent these features across
a range of physical parameters is tantamount to expecting them to fit
large, discontinuous, random signals.

However, by simply compensating for the inertial part of the measured
c.m. motion, the modulations can be almost completely eliminated (thick
transparent lines in the upper left panel).  It is notable that the
c.m. only drifts by roughly $0.1M$ during almost the entire inspiral
for the system shown in the upper panels of Fig.~\ref{fig:ampAndpos},
but still has such a drastic effect on the waveform's modes.  Even
though this is only a gauge choice---which we have been trained to
consider irrelevant in principle---in practice, gauge choices must be
made consistently and systematically for the waveforms to be really
useful.  In this sense, we might suggest that the c.m.-centered gauge
is really an \emph{optimal} choice.

A more difficult comparison is for the precessing system shown in the
lower panels of Fig.~\ref{fig:ampAndpos}.  The precession already
mixes the modes drastically, leading to a complicated waveform with
pronounced amplitude modulations, even after c.m. correction.  Clearly
the c.m. correction \emph{changed} the data, but it is not obvious that
we can say it was a change \emph{for the better}---at least from
looking at this plot alone.  To make the comparison more quantitative,
we introduce a new measure of a waveform's ``simplicity'' in
Sec.~\ref{sec:Accuracy}.  Essentially, this quantity measures the
residual when the waveform is modeled by simple linear-in-time
amplitudes in the corotating frame~\cite{PhysRevD.87.104006}.  The
value of this residual is 117 times smaller for the c.m.-corrected data
than for the raw data in this precessing system, showing that the
corrected waveform is clearly and objectively better in this sense at
least.

To address the miscalculation of the c.m. and its correction, this
paper is organized as follows:
\begin{enumerate}[i]
  \item{In Sec.~\ref{sec:COMneed}, we discuss the current definition of
    the c.m. and the consequences this definition and its use have on SXS
    gravitational waveforms.}
  \item{In Sec.~\ref{sec:COMCorrectionMethod}, we discuss the current
    method for correcting waveform data and selecting an optimal gauge.
    Any correlations found between simulation parameters and the c.m.
    correction factors are discussed.}
  \item{In Sec.~\ref{sec:Accuracy}, we discuss a quantitative method for
    evaluating the ``correctness'' of the gauge a waveform is currently
    in. We compare the waveform data in its original, unoptimized gauge
    to the c.m.-corrected gauge described in
    Sec.~\ref{sec:COMCorrectionMethod}.}
  \item{In Sec.~\ref{sec:ImprovingCOM}, we discuss how we may improve
    the definition of the c.m. to find a better correction, potentially
    leading to a further optimized choice of gauge. This section also
    investigates alternative definitions of the c.m. with a focus on
    potential physical causes of c.m. motion like that seen in
    Fig.~\ref{fig:ampAndpos}, including post-Newtonian definitions and
    considerations of linear-momentum recoil.}
  \item{Finally, we present our findings and results in
    Sec.~\ref{sec:Conclusions}.}
\end{enumerate}

\section{The need for c.m. corrections}
\label{sec:COMneed}

One of the primary concerns with BBH simulations with regards to
gravitational-wave astronomy is the validity of their gravitational
waveforms. Above all, the output from a BBH simulation should result
in a reliable, reproducible waveform that can then be released for
public usage. In the case of the SXS collaboration, many of the
waveforms produced are also compressed into a catalog that is released
to LIGO for data analysis and waveform comparisons with their
gravitational wave detector data.

Gravitational waveforms in the SXS catalog are given in terms of the
gravitational-wave strain $h$, or the Weyl component $\Psi_4$.  In
regards to detecting gravitational waves, $h$ and $\Psi_4$ contain the
same information, and the analysis and corrections applied in this
work may be applied to either with the same results. For simplicity,
we will focus on $h$.

Waveforms from SXS are represented by mode weights, or amplitudes, for
spin-weighted spherical harmonics (SWSHs).  The gravitational-wave
strain may be represented by the transverse-traceless projection of
the metric perturbation caused by the gravitational waves at time $t$
and location $(\theta, \phi)$ relative to the binary, and can be
combined into a single complex quantity, given by
\begin{equation}
  \label{h_decomp}
  h(t, \theta, \phi) \coloneqq 
  h_{+}(t, \theta, \phi) - \mathrm{i} h_{\times}(t, \theta, \phi) .
\end{equation}
For each slice in time, the combined perturbation $h(t, \theta, \phi)$
is measured on the coordinate sphere. The angular dependence of this
measurement can then be expanded in SWSHs.  The quantity
$h(t, \theta, \phi)$ has a spin weight of $-2$ \cite{Newman1966}, and
may be represented as
\begin{equation}
  \label{eq:h_YlmSum}
  h(t, \theta, \phi) = \sum_{l,m} h^{l,m}(t)\,{}_{-2}Y_{l,m}(\theta, \phi),
\end{equation}
where the complex quantities $h^{l,m}(t)$ are referred to as modes or
mode weights, and are much more convenient when analyzing BBH than the
total perturbation in any particular direction \cite{BoyleEtAl2014,
  Brown2007}.  Spin-weighted Spherical harmonics are further discussed 
in Appendix~\ref{sec:SWSH}.

The expansion in Eq.~(\ref{eq:h_YlmSum}) depends on orientation of the
spherical coordinates $\theta, \phi$ and their origin.  The customary
choice places the binary at the origin with the binary's initial
orbital plane coinciding with the equatorial plane $\theta=\pi/2$.
For comparable mass, nonprecessing binaries, the quadrupolar
$(l,m)=(2,\pm2)$ modes then dominate the waveform.  While the
$h^{2, \pm 2}$ modes are dominant, it is important to consider the
behavior of the other modes present in the waveform. The other modes
may not be used directly for BBH detection currently, as they are much
smaller in magnitude compared to the $h^{2, \pm 2}$ modes for most
systems, but can be important for parameter
estimation~\cite{Varma2017, Lange2017,
  Chatziioannou:2019dsz}. Additionally, there are proposals for using
higher-order modes in BBH searches~\cite{Pekowsky2012, Bustillo2016,
  Harry2018}.  Higher-order modes are also useful for verifying the
reliability and potentially the accuracy of the waveform. If the
shape, variability, magnitude, or any other characteristic of the
higher order, or subdominant, modes are found to not suitably match
with theory, then this could indicate a possible flaw in the
simulation.

One clear issue is the coordinate system, or gauge choice, for the
simulation, as spherical harmonics and hence SWSHs depend on the
coordinates. The center chosen for the simulation is the c.m. of the
system, as calculated and set in the initial data. It is expected that
the c.m. will move slightly throughout the simulation; however, large
movements are not expected and suggest a flaw in the choice of
gauge. If the c.m. moves significantly, there is mode
mixing~\cite{Boyle2016}. The dominant effect~\cite{Boyle2016} is
leaking of the $h^{2,\pm 2}$ modes of BBH waveforms into the higher
modes, and this leakage can be at least partially removed through c.m.
drift corrections, as described in Sec.~\ref{sec:COMCorrectionMethod}.

Mode mixing is manifested in the waveforms as oscillating amplitudes,
which can clearly be seen in the left column panels of
Fig.~\ref{fig:ampAndpos}, especially for the $(2,1)$, $(3,1)$, and
$(3,3)$ modes in the top panel for SXS:BBH:0314. Precessing
simulations, like SXS:BBH:0622, are expected to have some amplitude
modulations purely due to the orientation of the system.  The worse
the c.m. calculation is for a simulation, the more altered the SWSH
representations are, and the worse the mode mixing becomes.

It is easily seen in Fig.~\ref{fig:ampAndpos} that the applied c.m.
correction removes what we will find to be unphysical waveform
amplitude modulations for nonprecessing systems, however for
precessing systems it is not so clear or obvious. Therefore,
especially for precessing systems, a quantitative method is required
for evaluating the ``correctness'' of the c.m. and gauge.  This is
discussed in Sec.~\ref{sec:Accuracy}.

The current definition used during BBH simulations for the c.m. is the
usual Newtonian definition:
\begin{equation}
  \label{eqn:comdef}
  \vec{x}_{\text{c.m.}} = \frac{m_{a}}{M} \vec{x_a} + \frac{m_{b}}{M} \vec{x_b},
\end{equation}
where $M = m_a + m_b$ is the total mass of the system, $m_a$ and $m_b$
are the Christodoulou masses~\cite{christodoulou1970} of the primary
and secondary black holes respectively, and $\vec{x}_a$ and
$\vec{x}_b$ are the coordinates of the centers for the primary and
secondary black hole respectively.  This is a Newtonian expression for
the c.m., and from output of the simulations like in
Fig.~\ref{fig:ampAndpos}, we know is not a perfect description of the
optimal c.m.. The tracking of the c.m. throughout the simulations can be
seen for SXS:BBH:0314 and SXS:BBH:0622 in the right column panels of
Fig.~\ref{fig:ampAndpos}.

The c.m. motion is an effect of the initial data.  One aspect of the
initial data construction method proposed in Ref.~\cite{Ossokine2015}
is the elimination of Arnowitt-Deser-Misner (ADM), or spatial, linear 
momentum in the initial data for
precessing systems, namely enforcing $\vec{P}_{\mathrm{ADM}}=0$.  The
work done in Ref.~\cite{Ossokine2015} proposed a new, and now adopted,
method for calculating and constructing the initial data for BBH
simulations. The improved method for calculating initial data has
far-reaching effects in Spectral Einstein (SpEC) simulations and 
most of the simulations
in the SXS simulation catalog were completed using this relatively new
method. This had the effect of reducing specific components of mode
mixing as seen in the gravitational waveforms, however as showcased in
Fig.~\ref{fig:ampAndpos}, significant mode mixing is still present.

As is further discussed in Sec.~\ref{sec:linearmom}, linear-momentum
recoil is an expected physical contribution to the motion of the
c.m.. However, unphysical contributions to the linear momentum in the
initial data of simulations introduce unphysical motion in the c.m.,
essentially imparting spurious linear-momentum kicks. By controlling
the linear momentum and removing it, this effect from the initial data
is removed. However, even for simulations with initial data
constructed using the method described in Ref.~\cite{Ossokine2015},
significant translations and boosts, and the resulting mode mixing,
are still present in the gravitational waveforms. This warrants
further investigation into the c.m. motion and the application of a c.m.
correction.

It had been suggested in Refs.~\cite{Boyle2016, Ossokine2015} that 
much of the c.m. motion depicted in the
right column panels of Fig.~\ref{fig:ampAndpos}, and seen in all SXS
simulations, was largely unphysical and could be removed from the
data. The c.m. correction used to remedy the unphysical c.m. motion is
discussed in the following section. Additionally, there are
alternative definitions of the c.m. and physical effects that are
expected to cause the c.m. to move, or imply that the c.m. is not moving
at all. The more obvious of these physical effects are post-Newtonian
(PN) corrections for the c.m. which may include effects explaining the
c.m. motion, and linear-momentum recoil from the system. PN and linear
momentum contributions are examined in Secs.~\ref{sec:PN}
and~\ref{sec:linearmom}.

\section{Centre-of-mass correction method}
\label{sec:COMCorrectionMethod}

Previous work~\cite{Boyle2016, Ossokine2015} suggests that the c.m.
motion is largely a result of gauge choice.  Therefore, understanding
the c.m. correction begins with understanding the permissible gauge
transformations.  More specifically, we are interested in the gauge
transformations that will affect the waveform measured by distant
observers.  Because a gravitational-wave detector will typically be
very far from the source, only the asymptotic behavior of the waves is
generally considered relevant---specifically at future null infinity,
$\scriplus$.

While the asymptotic gauge of waveforms from numerical relativity has
not been extensively investigated, it is certainly fair to say that no
results in the literature thus far have been in a completely specified
gauge.  Even the strongest claims of ``gauge-invariant'' asymptotic
waveforms~\cite{HandmerEtAl2015} are only invariant modulo the
\emph{infinite-dimensional} Bondi-Metzner-Sachs (BMS) gauge
group~\cite{Bondi:1962px, Sachs:1962wk}, which is the asymptotic
symmetry group corresponding to the Bondi gauge condition.  An
important feature of Bondi gauge is that the gravitational waves
measured by \emph{any} distant inertial observer (at least over a
duration short compared to the distance to the source) are
approximately given by the asymptotic metric perturbation at fixed
spatial coordinates as a function of retarded time in some member of
this gauge class---and conversely, any such function corresponds to a
signal that could be measured by some distant inertial
observer~\cite{StromingerLectures}.  Essentially, we might think of
Bondi gauge to be ``as simple as possible, but not simpler'' for the
purposes of gravitational-wave detection.  Because the BMS group
alters the waveform while preserving Bondi gauge, we consider it to be
the fundamental symmetry group relevant to gravitational-wave
modeling.\footnote{Other possible gauge choices exist.  For example,
  Newman-Unti gauge~\cite{1962JMP.....3..891N} is closely related to
  Bondi gauge, and is invariant under the same asymptotic symmetry
  algebra~\cite{BarnichLambert}.  More generally, it is not even clear
  that waveforms from numerical relativity are actually expressed in
  either of these well-defined gauge classes, in which case more
  general gauge transformations may be of interest.  Ultimately, the
  gauge freedoms relevant to counteracting c.m. motion are simply space
  translations and boosts.  As long as these transformations are
  allowed, this discussion of c.m. motion remains relevant.  Previous
  work~\cite{PhysRevD.88.124010} suggests that SXS waveforms are
  consistent with waveforms in Bondi gauge, though further research is
  warranted.}

Because BMS transformations preserve the inertial property of
observers, we cannot expect to counteract all of the c.m. motion seen
in Fig.~\ref{fig:ampAndpos}---particularly the cyclical behavior.
However, in addition to the cyclical behavior, these coordinate tracks
begin with some overall displacement from the origin, and then drift
away from that initial location over the entire course of the
inspiral.  Thus, we expect that a space translation and a boost are
needed to negate the effects of some of the c.m. motion.  In
particular, we will choose the translation $\vec{\alpha}$ and boost
$\vec{\beta}$ to minimize the average of the square of the distance
between the measured c.m. and the origin.\footnote{This measure will be
  invariant under time translation and rotation, which are generally
  dealt with separately during gravitational-wave analysis, so we
  simply ignore those degrees of gauge freedom.  Furthermore, it is
  not at all clear how a higher-order supertranslation should affect
  the coordinates close to the center of a simulation, and so we leave
  discussion of more general supertranslations to future work.}

\begin{figure*}
  \includegraphics[width=\textwidth]{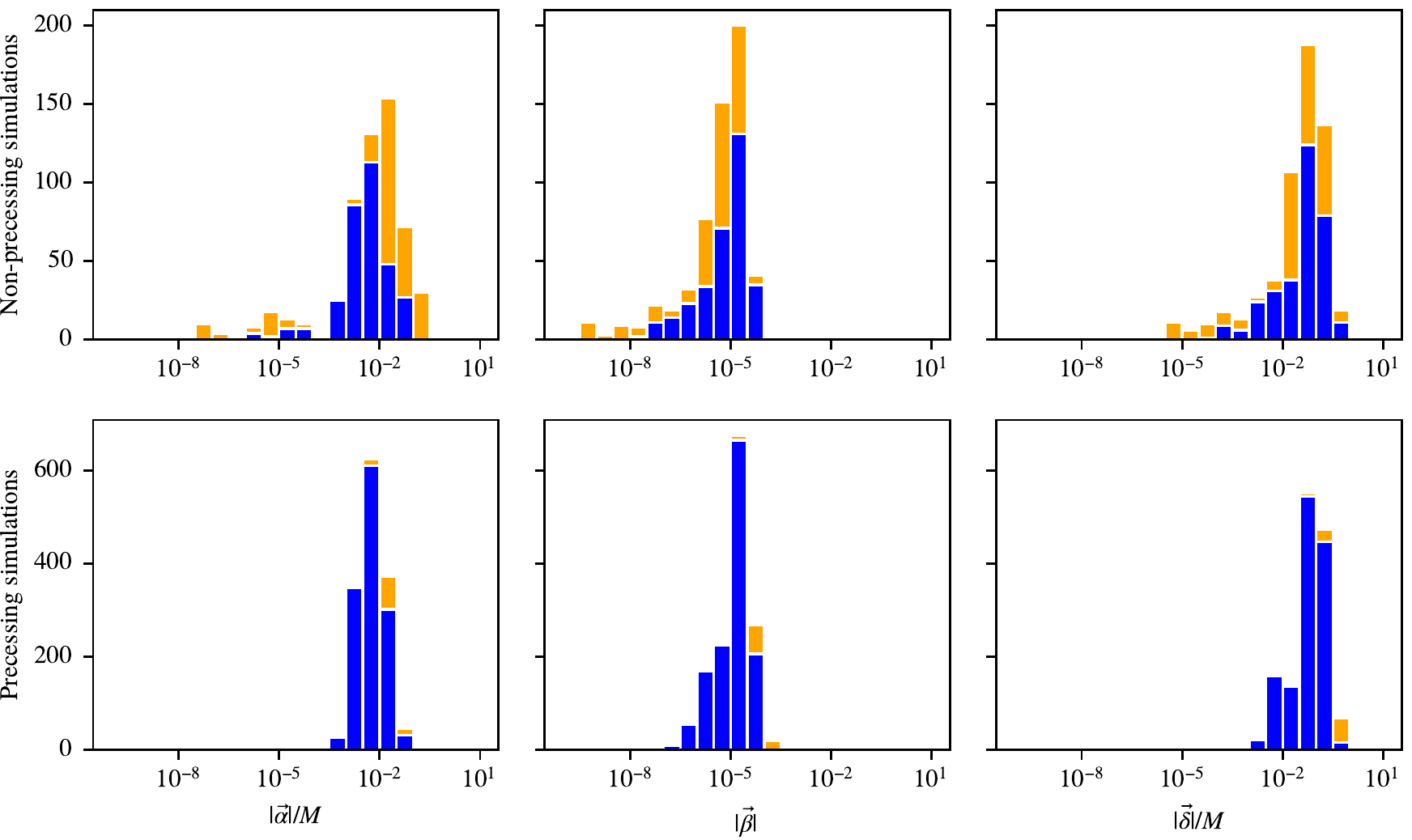}
  \caption{\label{fig:c.m.valueoverview}%
    Magnitudes of c.m. offsets and drifts for all simulations in the
    SXS catalog. The top row shows values for nonprecessing systems
    (i.e., nonspinning, spin aligned, and spin antialigned) and the
    bottom row shows values for precessing systems. The horizontal
    axis for each plot is the magnitude of the c.m. value shown
    ($|\vec{\alpha}|, |\vec{\beta}|,$ or
    $|\vec{\delta}| = |\vec{\alpha} + \vec{\beta}t_{\text{merger}}|$,
    where $t_{\text{merger}}$ is the first reported instance of a
    common apparent horizon found between the two BHs) and the
    vertical axis is the number of simulations that have c.m. values of
    that bin magnitude.  Note that typical values of $|\vec{\beta}|$
    are quite small, but accumulate over the course of a simulation to
    cause a large overall displacement by merger. Blue indicates runs
    using the initial-data method described in
    Ref.~\cite{Ossokine2015}; orange indicates runs using the previous
    initial-data method.  These results suggest that this procedure
    improves the initial location of the center of mass, but does
    little to improve its drift.}
\end{figure*}

\subsection{Choosing the translation and boost}
\label{sec:choos-transl-boost}

We follow Appendix~E of Ref.~\cite{Boyle2016} in choosing the
translation $\vec{\alpha}$ and boost $\vec{\beta}$ to minimize the
average square of the distance between the c.m. measured in the raw
data and the origin of the corrected frame.  That is, we choose
$\vec{\alpha}$ and $\vec{\beta}$ to minimize the function
\begin{equation}
  \label{eqn:Xi}
  \Xi(\vec{\alpha},\vec{\beta}) = \int_{t_i}^{t_f}|\vec{x}_{\mathrm{c.m.}}
  - (\vec{\alpha}+\vec{\beta}t)|^2dt.
\end{equation}
It is not hard to find the minimum of this quantity analytically.  We
define two moments of the c.m. position
\begin{subequations}
  \label{eqn:com_moments}
  \begin{gather}
    \label{eqn:com_moment1}
    \langle\vec{x}_{\text{c.m.}}\rangle =
    \frac{1}{t_f-t_i}\int_{t_i}^{t_f} \vec{x}_{\text{c.m.}}(t)\, dt,
    \\
    \label{eqn:com_moment2}
    \langle t\vec{x}_{\text{c.m.}}\rangle =
    \frac{1}{t_f-t_i}\int_{t_i}^{t_f} t\, \vec{x}_{\text{c.m.}}(t)\, dt.
  \end{gather}
\end{subequations}
Then, the minimum of Eq.~\eqref{eqn:Xi} is achieved with
\begin{subequations}
  \label{eqn:dx_v_def}
  \begin{gather}
    \vec{\alpha} = \frac{4(t_f^2+t_ft_i+t_i^2)\langle\vec{x}_{\text{c.m.}}\rangle - 
      6(t_f+t_i)\langle t\, \vec{x}_{\text{c.m.}}\rangle
    }{(t_f-t_i)^2}, \\
    \vec{\beta} = \frac{12\langle t\, \vec{x}_{\text{c.m.}}\rangle - 
      6(t_f+t_i)\langle\vec{x}_{\text{c.m.}}\rangle}{(t_f-t_i)^2}.
  \end{gather}
\end{subequations}
We then apply this transformation to the asymptotic waveform using the
method described in Ref.~\cite{Boyle2016}.  Note that this rests on
implicit assumptions about how directly comparable the coordinates of
the apparent-horizon data and the asymptotic coordinates are.  For
example, this assumes that the time coordinate of the apparent-horizon
data and the asymptotic retarded-time coordinate are equal.  While
there is no rigorous motivation for this assumption, the results of
Sec.~\ref{sec:Accuracy} bear out its approximate validity.

Using this minimization method, the c.m. offsets for every public
waveform in the SXS catalog have been corrected in the waveform
data. The first instance of the c.m. corrections to waveforms in the
SXS public waveform catalog was in January of 2017. Center-of-mass corrected
waveform data is recommended over non corrected data in all cases, and
corresponding files are listed in the SXS public waveform catalog as
files ending in \verb|CoM.h5|. An overview of the c.m. correction
values is shown in Fig.~\ref{fig:c.m.valueoverview}.

It is clear from the upper-left panel of Fig.~\ref{fig:ampAndpos} that
c.m. removal is ``helpful'' in the sense that it reduces the amplitude
oscillations, which are not expected on physical grounds.
Unfortunately, this by-eye analysis is not quantitative, and it is not
clear how it would apply to a precessing system, as seen in the
lower-left panel of the same figure.  We discuss a better measure of
how the waveform quality is impacted by c.m. corrections in
Sec.~\ref{sec:Accuracy}.

\subsection{Choosing the integration region}
\label{sec:titf}

\begin{figure}
  \includegraphics[width=\columnwidth]{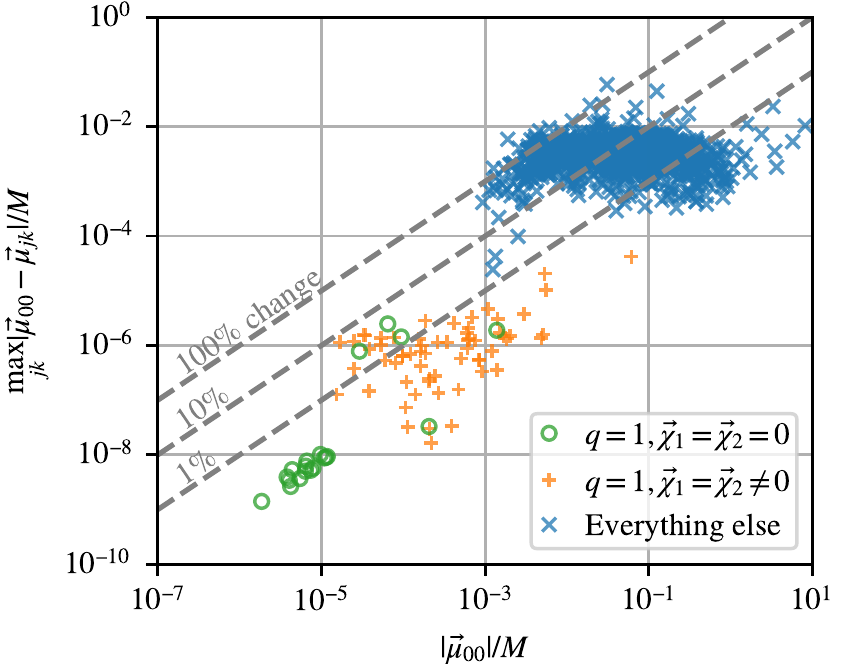}
  \caption{\label{fig:titferror}%
    Comparing the size of c.m. corrections in the SXS catalog,
    $|\vec{\mu}_{00}|$, to how much those corrections change under
    small variations in the end points of integration used to compute
    the c.m. correction.  The vertical axis shows the largest change in
    the c.m. correction if we shift $t_{i}$ and/or $t_{f}$ later by
    half an orbit.  The systems with the largest c.m.
    corrections---where these corrections are presumably the most
    important---change by small fractional amounts.  On the other
    hand, there are several systems in which the c.m. correction
    changes by more than the original correction; those systems also
    have some of the smaller c.m. corrections in the catalog.  The
    median percentage change is 4\% of the original correction, and
    even the largest individual change is smaller than the median
    value of $|\vec{\mu}_{00}|$.  }
\end{figure}

The determination of $\vec{\alpha}$ and $\vec{\beta}$ is made over a
subset of the total simulation time, from $t_i$ to $t_f$ [see
Eq.~\eqref{eqn:Xi}].  Choosing different values of $t_i$ and $t_f$ may
affect the resulting $\vec{\alpha}$ and $\vec{\beta}$ values.  For the
corrections performed on the SXS catalog, a standard subset of the
simulation time was chosen.  All waveforms had their c.m.-correction
values calculated from $t_{i} = t_{\text{relax}}$, the ``relaxation''
time after which the initial transients have dissipated, to 10\%
before the end of the inspiral: $t_{f} = 0.9 t_{\text{merger}}$. These
time bounds were chosen to avoid including periods of junk radiation
as well as the merger and ringdown stages.

However, changing $t_i$, $t_f$ by small amounts could change the c.m.
correction values. As there is epicyclical motion of the c.m. (as seen
in Fig.~\ref{fig:ampAndpos}, for example), changing the beginning or
ending time may cause the resulting $\vec{\alpha}$ and $\vec{\beta}$
to change, depending on where $t_i$ and $t_f$ fall on an epicycle. For
example, if $t_i$ and $t_f$ are separated by an integer number of
epicycles, then we might expect any effect from the epicycles on the
calculation of $\vec{\alpha}$ and $\vec{\beta}$ to cancel
out. However, if $t_i$ and $t_f$ are separated by a noninteger number
of epicycles, especially by a half-integer number of epicycles, the
epicycles may induce significant bias in $\vec{\alpha}$ and
$\vec{\beta}$.  The overall number of epicycles included in the
calculation of $\vec{\alpha}$ and $\vec{\beta}$ may also affect how
sensitive they are to this bias.

Here, we compare the size of the c.m. correction using the standard
prescription to the size of the correction when $t_i$ and/or $t_f$ are
changed by half an orbit.  This will give us some idea of the
stability of the c.m.-correction procedure.  However, it must be noted
that, at a larger scale, the choices of $t_{i}$ and $t_{f}$ are quite
arbitrary.  For some purposes, it may be preferable to choose those
values to range over only the first half of the inspiral, or even just
the ringdown stage.  The values used in the SXS catalog were chosen
for robustness and easy reproducibility.

To simplify the comparison, we describe the c.m. motion using the
quantity $\vec{\mu}$, which gives the most distant position of the
corrected origin of coordinates throughout the inspiral, relative to
the origin used in the simulation.  Specifically, we can define
$\vec{\mu}$ according to
\begin{equation}
  \label{eq:mu_definition}
  \vec{\mu} = \begin{cases}
    \vec{\alpha} & \text{if } |\vec{\alpha}|
    > |\vec{\alpha}+\vec{\beta}t_{\text{merger}}|, \\
    \vec{\alpha}+\vec{\beta}t_{\text{merger}} & \text{if }
    |\vec{\alpha}| \leq |\vec{\alpha}+\vec{\beta}t_{\text{merger}}|.
  \end{cases}
\end{equation}
For 96\% of the simulations in the SXS catalog, we find that
$\vec{\mu} = \vec{\alpha}+\vec{\beta}t_{\text{merger}}$.  The 4\% of
simulations with $\vec{\mu}=\vec{\alpha}$ have no apparent
correlations with system parameters, and are effectively random. We
also introduce subscripts, so that $\vec{\mu}_{00}$ is the result of
this calculation when using the original values of $t_{i}$ and
$t_{f}$; $\vec{\mu}_{10}$ is the result when moving $t_{i}$ later by
half an orbit; $\vec{\mu}_{01}$ is the result when moving $t_{f}$
later by half an orbit; and $\vec{\mu}_{11}$ is the result when moving
both $t_{i}$ and $t_{f}$ later by half an orbit.

In Fig.~\ref{fig:titferror}, we examine
$\max_{jk} |\vec{\mu}_{00} - \vec{\mu}_{jk}|$ as a measure of how
robust the c.m. corrections are with respect to these small adjustments
in the choices of $t_{i}$ and $t_{f}$.  In the great majority of
systems the c.m. changes by less than $10^{-2}M$.  This is, for
example, just one tenth the size of the displacements seen in the
upper panels of Fig.~\ref{fig:ampAndpos}.  The median change is
$3.1 \times 10^{-3}M$, and in all cases is smaller than the median
value of $|\vec{\mu}_{00}|$ itself, which is $6.9 \times 10^{-2}M$.
The systems with the largest c.m. corrections in the SXS catalog change
by fractions of a percent, suggesting that the results are certainly
stable in the cases where applying a c.m. correction is most important.
There are several cases where the fractional change is greater than
100\%, though these are systems with relatively small values of
$\vec{\mu}_{00}$.  The median fractional change is $4.3\%$.  It is
also notable that the data points separate roughly into three groups.
The group in the lower left corner of Fig.~\ref{fig:titferror} is
comprised exclusively of equal-mass nonspinning simulations with
various eccentricities, though several of these are also found in the
central group.  The central group is where all equal-mass simulations
with equal but nonzero spins are found, which includes ten systems
with significant precession.  Every other type of system is in or near
the largest group, on the upper right.

\begin{figure*}
  \includegraphics[width=\textwidth]{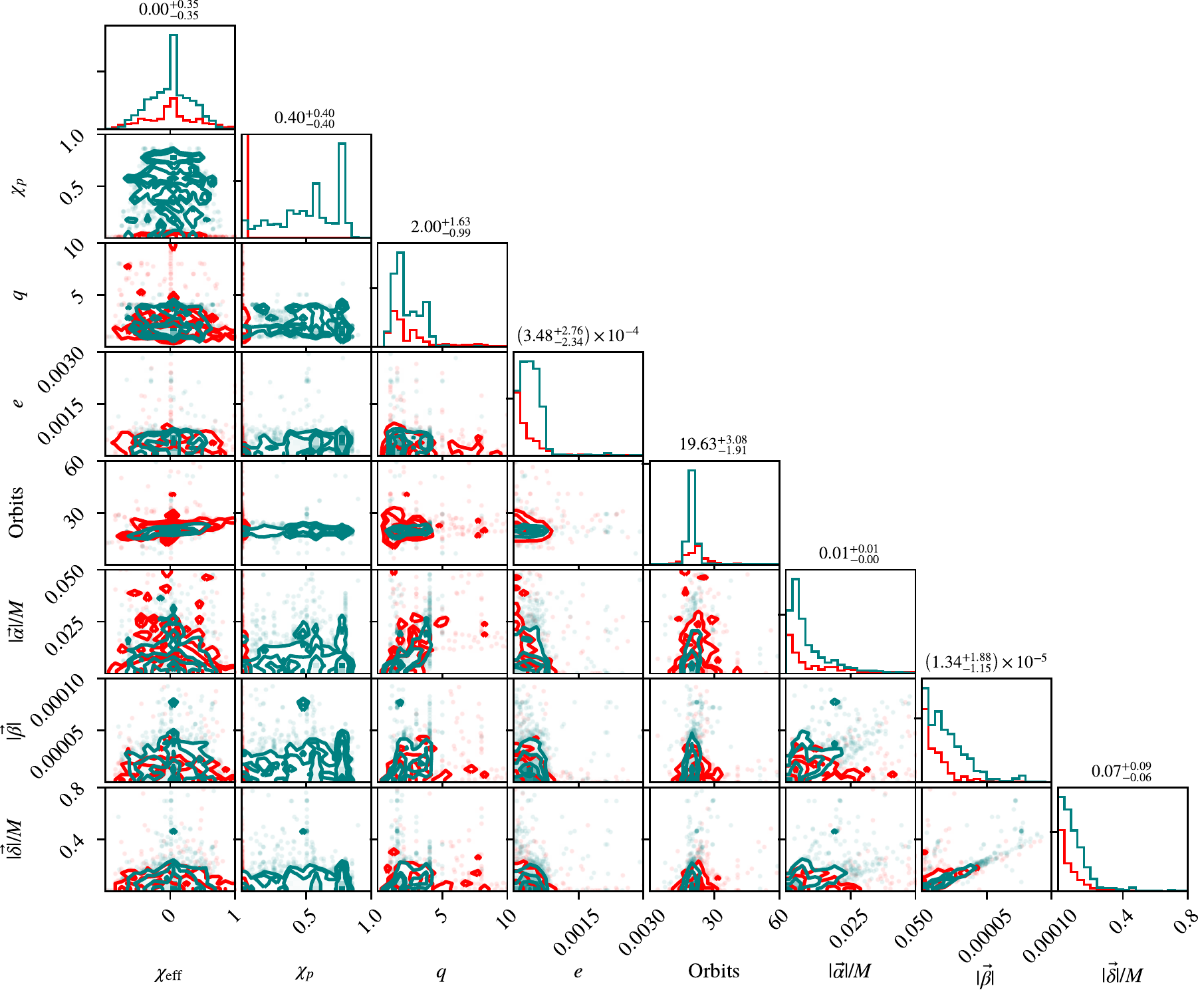}
  \caption{\label{fig:corner}%
    Conter-of-mass correction values and relevant simulation parameters.
    $\chi_{\text{eff}}$ is the effective spin, $\chi_p$ is the
    effective precessing spin, $q$ is the mass ratio, $e$ is the
    eccentricity, Orbits represents the total number of orbits the
    simulation had at $t_{\text{merger}}$, and $\vec{\alpha}$,
    $\vec{\beta}$,
    $\vec{\delta} = \vec{\alpha} + \vec{\beta}t_{\text{merger}}$ are
    the c.m. correction values representing the spatial translation,
    boost, and total c.m. displacement respectively. Red represents
    spin-aligned simulations, and teal represents precessing
    simulations.  The numbers above each column represent the median
    of each variable over all simulations, with superscripts and
    subscripts giving the offset (relative to the median) of the
    84th and 16th percentiles, respectively.}
\end{figure*}

\subsection{Correlations between c.m. correction values and physical
  parameters}
\label{sec:Correlations}

Along with having c.m. corrected the waveforms, we have also performed
an analysis of the values of the boosts and translations needed by
each simulation in the SXS public waveform catalog.

No obvious correlations can be seen in Fig.~\ref{fig:c.m.valueoverview}
between spin aligned and precessing systems. We also show a more
in-depth correlation plot in Fig.~\ref{fig:corner}, taking more of the
simulation parameters into consideration. It can be seen that
typically precessing simulations may have larger overall c.m.
displacement,
$\vec{\delta} = \vec{\alpha} + \vec{\beta}t_{\text{merger}}$, and that
larger boost values $\vec{\beta}$ correspond with larger overall
displacement values $\vec{\delta}$ for both spin aligned and
precessing systems.

Outside of the correlations between the boost $\vec{\beta}$ and total
displacement $\vec{\delta}$ of the c.m., there does not appear to be
any other strong correlations present for the current SXS simulation
catalog. It was expected that precessing, high mass ratio, and
eccentric systems should have vastly different c.m. correction values
than spin aligned, low mass ratio, and more circular systems, however
no such correlations are present with this data set.

For Fig.~\ref{fig:corner}, we use the eccentricity $e$, number of
orbits, and mass ratio $q$ reported by SpEC at the end of the
simulation.  We calculate the effective spin~\cite{Ajith:2009bn,
  Santamaria:2010yb, LIGO2016}
\begin{equation}
  \label{eqn:chi_eff}
  \chi_{\text{eff}}
  = \frac{c}{GM} \left( \frac{\vec{S}_a}{m_a} + \frac{\vec{S}_b}{m_b}
  \right) \cdot \frac{\vec{L}}{|\vec{L}|}
  = \frac{\vec{\chi}_am_a+\vec{\chi}_bm_b}{M}
  \cdot \frac{\vec{L}}{|\vec{L}|},
\end{equation}
and an effective precession parameter~\cite{Schmidt2015, LIGO2016}
\begin{equation}
  \label{eqn:chi_p}
  \chi_p
  = \frac{c}{B_1Gm_a^2}
  \max(B_1|\vec{S}_{a,\perp}|, B_2|\vec{S}_{b,\perp}|).
\end{equation}
Here, $M=m_a+m_b$ is the total mass of the system,
$\vec{S_i} = G/c\,\vec{\chi}_im_i^2$ is the angular momentum of the
$i$-th black hole and $\vec{\chi}_i$ its dimensionless spin,
$B_1 = 2+3m_b/2m_a$, $B_2 = 2+3m_a/2m_b$, and the subscript $\perp$
indicates the quantity perpendicular to the orbital angular momentum
$\vec{L}$, e.g.,
$\vec{S}_{a,\perp} = \vec{S}_a - (\vec{S}_a\cdot\hat{L})\hat{L}$.
Note that $\chi_p$ gives a measure of how much a system is precessing
during a simulation.

\section{Quantifying c.m. Correction Using Waveforms Alone}
\label{sec:Accuracy}

Any discussion of c.m. based on the positions of the individual black
holes will suffer from the same fundamental ambiguity: reliance on
coordinates---specifically in the highly dynamical region between the
two black holes---that are subject to unknown gauge ambiguities.  The
only region of the spacetime where the gauge freedom is limited in any
useful sense is the asymptotic region, in which we assume the only
freedom is given by the BMS group (described in
Sec.~\ref{sec:COMCorrectionMethod}).  While there are many suggestions
in the literature~\cite{1966JMP.....7..863N, 2010CQGra..27x5004A,
  2012LRR....15....1A, 2009CQGra..26f5008D, 2008CQGra..25n5001K,
  2004LRR.....7....4S, 2007GReGr..39.2125H, 1984CQGra...1...15D,
  1982JMP....23.2410A, 2016PhRvD..93j4007F} for using asymptotic
information to specify the asymptotic gauge more narrowly, they all
require more information than is available from most catalogs of
numerical-relativity waveforms---such as additional Newman-Penrose
quantities or more precise characterization of the asymptotic behavior
of the various fields.

Here, we present a simplistic but effective measure of c.m. effects
that can be applied exclusively to asymptotic waveform data $h$ or
$\Psi_{4}$.  The basic idea is that we expect to be able to decompose
a waveform measured in c.m.-centered coordinates into modes that are,
at least for small portions of the inspiral, given by a slowly
changing complex amplitude times a complex phase that varies
proportionally with the orbital phase.  When the waveform is
decomposed in off-center coordinates, those well-behaved modes mix, so
that the amplitude and phase do not behave as expected.  Therefore, we
will attempt to model a given waveform in a sort of piecewise fashion
that \emph{assumes} the expected behavior, and simply measure the
residual between the model and the waveform itself.  For a given
transformation applied to the waveform, we will minimize the residual
by adjusting the parameters to the model while keeping the waveform
fixed.  The smaller the minimized residual, the more accurately the
waveform with that transformation can be modeled in this simple way,
and the more nearly we expect that the waveform is decomposed in
c.m.-centered coordinates.  Roughly speaking, we can think of this as a
measure of the ``simplicity'' of the waveform, which is not only in
line with our basic expectations for waveforms in the appropriate
coordinates, but also a useful measure of how accurately simple
waveform models (EOB, surrogate, etc.) will be able to capture
features in the numerical waveforms.  This criterion is obviously
totally distinct from any criteria involving the BH positions, but is
important precisely because it provides a complementary way of looking
at the data.  Finding agreement between the results of this method and
another will lend support to the idea that the other method is
suitable.

\subsection{Defining the method}
\label{sec:defining-method}

We now describe this method more precisely.  The initial inputs are
some translation $\vec{\alpha}$ and boost $\vec{\beta}$ that we wish
to evaluate.  We transform the waveform by those inputs and denote the
result $\mathcal{T}_{\vec{\alpha}, \vec{\beta}} [ h ]$.  We then
transform to a ``corotating frame'', which is a time-dependent frame
chosen so that the waveform in that frame is varying as slowly as
possible~\cite{PhysRevD.87.104006}.  Only the angular velocity of this
frame, $\vec{\Omega}$, is determined by the condition that the
waveform vary slowly; it is integrated in time to obtain one such
frame~\cite{Boyle2017}, but the result is only unique up to an overall
rotation.  We choose that overall rotation so that the $\vec{z}\,'$
axis of the final corotating frame is aligned as nearly as possible
throughout the inspiral portion of the waveform with the dominant
eigenvector~\cite{PhysRevD.84.124002, PhysRevD.84.124011} of the
matrix
\begin{equation}
  \langle L_{(a}L_{b)} \rangle \defined \int_{S^{2}}
  L_{(a}\left\{\mathcal{T}_{\vec{\alpha}, \vec{\beta}} [h] \right\}^\ast
  L_{b)}\left\{\mathcal{T}_{\vec{\alpha}, \vec{\beta}} [h] \right\}\,
  dA,
\end{equation}
where $L_{a}$ is the usual angular-momentum operator.  This still
leaves the frame defined only up to an overall rotation about
$\vec{z}\,'$, but such a rotation will have no effect on our results.
The transformed waveform in this corotating frame will be denoted
$\mathcal{R} \left\{ \mathcal{T}_{\vec{\alpha}, \vec{\beta}} \left[ h
  \right] \right\} (t, \theta, \phi)$, though we will usually suppress
the parameters, and may decompose the angular dependence in terms of
SWSH mode weights as usual.  This is the quantity we will be
attempting to model.

For the model itself, we first break the inspiral up into smaller
spans of time; we will be modeling the waveforms using simple
linear-in-time approximations, so we cannot expect to accurately
reproduce the nonlinear evolution over a very long portion of the
inspiral using just one such model.  The relevant measurement of the
waveform's dynamical behavior is the angular velocity of the
corotating frame.  More specifically, we define
$\Omega_{z'} = \vec{\Omega} \cdot \vec{z}\, '$, and use that to
determine a phase\footnote{This phase is loosely related to the
  orbital phase of the binary.  The angular velocity $\vec{\Omega}$,
  however, is defined solely with respect to the waveform at
  $\scriplus$, and entirely without reference to any quantities at
  finite distance in the system.  Nonetheless, for reasonably
  well-behaved coordinate systems, we would expect it to agree roughly
  with the orbital phase deduced from the trajectories of the black
  holes, especially during the early inspiral regime.}
$\Phi_{z'} = \int \Omega_{z'}\, dt$.  An obvious span of time would be
a single cycle of this phase, which would include enough data so that
the fit would actually reflect the behavior of the waveform, but not
so much that we would expect a poor fit due to evolution on the
inspiral timescale.  However, we will essentially be fitting
oscillatory terms with linear models.  In the simple case of fitting a
line to a basic sine function, it is not hard to see that the optimal
line has the expected slope of zero---independent of the phase of the
sine function---when the fit region is such that the argument of the
sine function goes through a phase change of $\varphi \approx 8.9868$
[or other solutions of $\varphi = 2 \tan(\varphi / 2)$].  Therefore,
we select each span of time so that it extends over a phase
$\Phi_{z'}$ of approximately $\varphi$, thereby determining the
difference in time between $t_{i,1}$ and $t_{i,2}$ so that they
satisfy
\begin{equation}
  \label{eq:time_spans}
  \Phi_{z'}(t_{i,2}) - \Phi_{z'}(t_{i,1}) = \varphi.
\end{equation}
We find that this choice does drastically reduce the oscillations in
the optimal fit parameters as we shift the fitting window.  While the
individual time spans extend over this range, we find that remaining
effects from oscillation are minimized by selecting successive time
spans to be separated by \emph{half} of a period---so that $\Phi_{z'}$
changes by exactly $\pi$ between $t_{i,1}$ and $t_{i+1,1}$:
\begin{equation}
  \label{eq:successive_times}
  \Phi_{z'}(t_{i+1,1}) - \Phi_{z'}(t_{i,1}) = \pi.
\end{equation}
So that the model may be reasonably accurate, without encountering
excessive numerical noise or excessively dynamical behavior at merger,
we limit the region over which we choose these time spans to be the
central 80\% of time between the ``relaxation time'' listed in the
waveform metadata and the time of maximum signal power in the
waveform.  This establishes $t_{0, 1}$, and all successive times can
be computed from that using Eqs.~\eqref{eq:time_spans}
and~\eqref{eq:successive_times}.

Now, we model the waveform ``piecewise'' on these spans of time,
though the pieces are overlapping.  The advantage of transforming the
waveform as described above is that each mode
separates~\cite{BoyleEtAl2014} into two parts that are symmetric and
antisymmetric under reflection along the $z$ axis.  The symmetric part
varies on an inspiral timescale because the primary rotational
behavior has been factored out by transforming to the corotating
frame; the antisymmetric part is mostly due to spin-orbit coupling and
therefore varies most rapidly by a complex phase with frequency equal
to the rotational frequency of the frame itself, though possibly with
opposite sign.  We model these two parts separately as simple
linear-in-time complex quantities, with an additional phase-evolution
term for the antisymmetric parts.  For each time span $i$, we write
\begin{multline}
  \label{eq:waveform_model}
  \mu_{i}(t, \theta, \phi)
  =
  \sum_{l, m} \left[
    s_{i}^{l, m} + \dot{s}_{i}^{l, m} (t - t_{i,c}) \right.
  \\ \left.
    + \left( a_{i}^{l, m} + \dot{a}_{i}^{l, m} (t - t_{i,c}) \right)
    \e^{i \sigma(m, l)\, \Phi_{z'}(t)}
  \right] {}_{-2}Y_{l,m}(\theta, \phi).
\end{multline}
Here, each of the coefficients $s_{i}^{l, m}$, $\dot{s}_{i}^{l, m}$,
$s_{i}^{l, m}$, and $\dot{a}_{i}^{l, m}$ is a complex constant, we use
$t_{i,c} = (t_{i,1} + t_{i,2})/2$ to mitigate degeneracy between the
constant-in-time and linear-in-time terms, and the function $\sigma$
is given by
\begin{equation}
  \label{eq:sign_function}
  \sigma(m, l) =
  \begin{cases}
    1 & |m| < l, \\
    -1 & |m| = l.
  \end{cases}
\end{equation}
These signs are chosen because they represent the dominant behavior of
the corresponding terms in the data.  Note that the symmetry
properties imply that once the quantities $s_{i}^{l, m}$, etc., are
chosen for positive $m$, they are automatically known for negative $m$
from the relations
\begin{subequations}
  \begin{gather}
    s_{i}^{l, m} = (-1)^{l} \bar{s}_{i}^{l, -m},
    \qquad
    \dot{s}_{i}^{l, m} = (-1)^{l} \bar{\dot{s}}_{i}^{l, -m},
    \\
    a_{i}^{l, m} = (-1)^{l+1} \bar{a}_{i}^{l, -m},
    \qquad
    \dot{a}_{i}^{l, m} = (-1)^{l+1} \bar{\dot{a}}_{i}^{l, -m}.
  \end{gather}
\end{subequations}
Because the $m=0$ modes of the SXS waveforms we use are not considered
reliable~\cite{PhysRevD.80.124045, SXSCatalog2019}, we simply ignore
those modes in both the model and the data.  That is, the sum in
Eq.~\eqref{eq:waveform_model} does not include any $m=0$ modes.  If
the sum over modes extends from $l=2$ to some maximum $l=L$, the total
number of (real) degrees of freedom in this model is $4 L(L+1) - 8$
for each span of time.  While the data we use contains up to $l=8$,
the highest-order modes contribute little to the result, and
drastically increase the number of degrees of freedom in the problem
(and therefore the time taken to optimize the model).  Therefore, we
use only up to $l=6$ in constructing the model and evaluating the
residual, reducing the degrees of freedom from 280 to 160 per time
span.  Finally, because this is still such a large number of degrees
of freedom, we limit the evaluation to only the first two and last two
time spans; we find that including the rest has no significant effect
on the result, but vastly increases the amount of processing time
required.  This leaves us with a manageable 640 degrees of freedom in
this model.

Now, using this model, we define the objective function
\begin{align}
  \Upsilon
  \left( \vec{\alpha}, \vec{\beta}
  \right)
  &=
    \min_{\boldsymbol{s}, \dot{\boldsymbol{s}},
    \boldsymbol{a}, \dot{\boldsymbol{a}} }
    \sum_{i} \int_{t_{i,1}}^{t_{i,2}} \int_{S^{2}}
    \left| \mathcal{R} \left\{ \mathcal{T}_{\vec{\alpha}, \vec{\beta}}
    \left[ h \right] \right\} - \mu_{i} \right|^{2}
    dA\, dt
    \nonumber \\
  \label{eq:objective_function}
  &=
    \min_{\boldsymbol{s}, \dot{\boldsymbol{s}},
    \boldsymbol{a}, \dot{\boldsymbol{a}} }
    \sum_{i} \int_{t_{i,1}}^{t_{i,2}} \sum_{l, m}
    \left| \mathcal{R} \left\{ \mathcal{T}_{\vec{\alpha}, \vec{\beta}}
    \left[ h \right] \right\}^{l, m} - \mu_{i}^{l,m} \right|^{2}
    \, dt.
\end{align}
We will use this function in two ways: first, to simply evaluate
$\Upsilon$ for given values of $(\vec{\alpha}, \vec{\beta})$, where
those values are obtained from the methods described in other
sections; second, to minimize $\Upsilon$ over possible values of
$(\vec{\alpha}, \vec{\beta})$ to find the optimum c.m. correction.

\subsection{Results for the standard c.m.-correction method}
\label{sec:results}

\begin{figure}
  \includegraphics[width=\columnwidth]{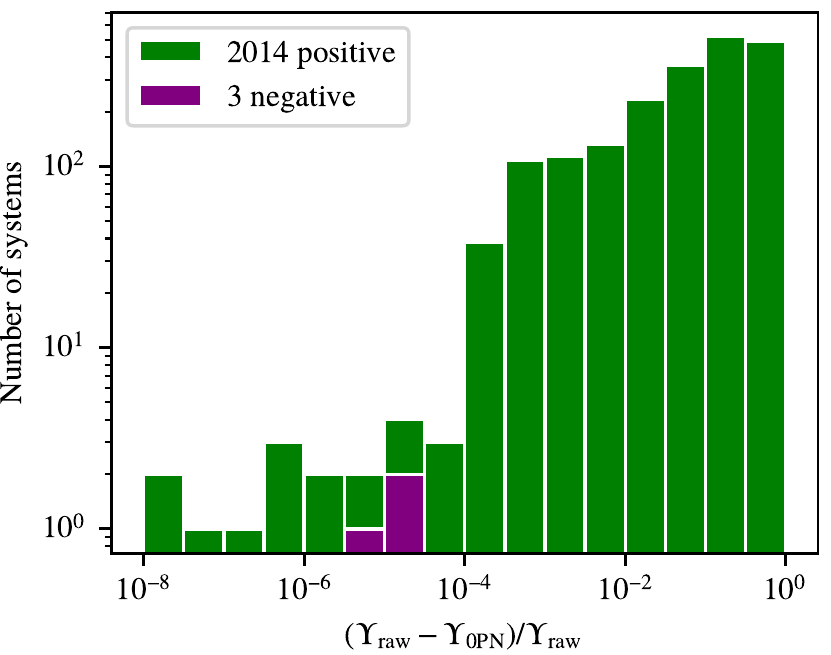}
  \caption{\label{fig:upsilon_histogram}%
    Relative difference between $\Upsilon_{\mathrm{raw}}$ evaluated on
    the raw waveform data and $\Upsilon_{\mathrm{0PN}}$ evaluated
    using the values $\vec{\alpha}$ and $\vec{\beta}$ given by the
    simplest Newtonian (0PN) approximation of
    Eqs.~\eqref{eqn:dx_v_def}---the same c.m. correction used in the
    current SXS catalog.  In the vast majority of cases, the value of
    $\Upsilon$ decreases substantially (though it actually
    \emph{increases} very slightly in three cases with significant
    eccentricity).  This suggests that even though $\Upsilon$ and the
    coordinate-based c.m. are such entirely different measures and
    based on completely different data, they agree that the changes
    introduced by naive c.m. corrections are generally improvements.  }
\end{figure}

We can now compare the value of $\Upsilon$ defined in
Eq.~\eqref{eq:objective_function} for all the waveforms discussed in
the previous sections.  First, we compare its value
$\Upsilon_{\mathrm{raw}}$ in the raw data to its value
$\Upsilon_{\mathrm{0PN}}$ using $\vec{\alpha}$ and $\vec{\beta}$ as
given by Eqs.~\eqref{eqn:dx_v_def}, where $\vec{x}_{\mathrm{c.m.}}$ is
given by the Newtonian (0PN) formula.  The latter corresponds to the
technique actually used in the current SXS data, for waveforms found
in the SXS simulation catalog with file names ending in \texttt{CoM}.
The results of this comparison are shown in
Fig.~\ref{fig:upsilon_histogram}.  One unusually short simulation
(SXS:BBH:1145~\cite{SXS:BBH:1145}) in the SXS catalog did not have
enough GW cycles to evaluate $\Upsilon$ properly, leaving a total of
2,017 systems shown in these figures.  The vast majority of systems
improve significantly by this measure.  The notable feature is that
even though the naive 0PN c.m. trajectory is so fundamentally different
from $\Upsilon$, this plot suggests that they agree in the sense that
the 0PN correction improves the waveforms for all but three
systems---and even for those three the change is very
small.\footnote{These three systems are unusual, in that they are
  quite short (having 13 to 15 orbits before merger, compared to an
  average of 22), and have eccentricities (0.215 and higher) that
  place them among the 12 most eccentric in the SXS catalog.
  Furthermore, the magnitude of the change in $\Upsilon$ for each of
  them is very small---in the lowest percentile for the entire
  catalog---which suggests that the negative results may be consistent
  with numerical error, and in any case are not cause for much
  concern.}

\begin{figure}
  \includegraphics[width=\columnwidth]{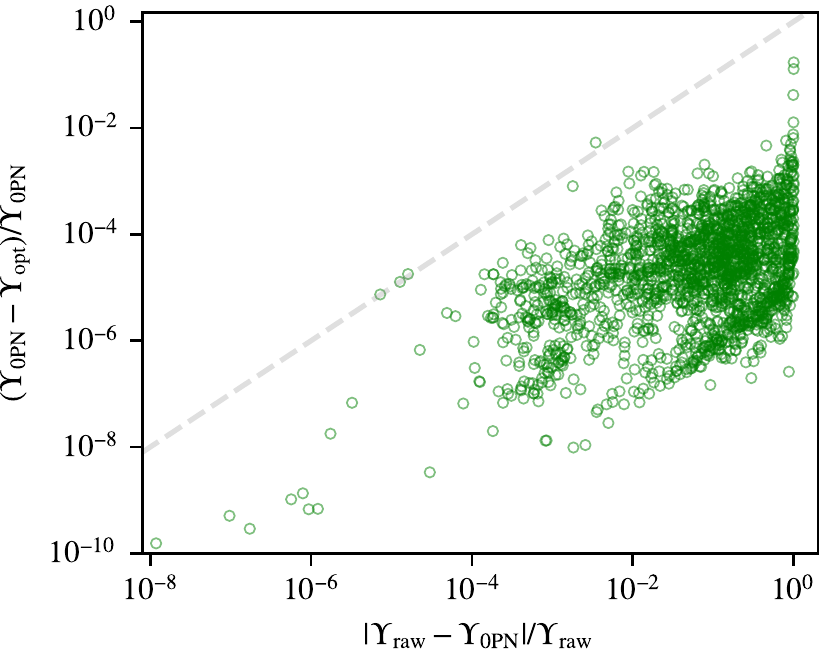}%
  \caption{\label{fig:upsilon_comparison_opt}%
    Comparison between the value $\Upsilon_{\mathrm{opt}}$ for which
    $\vec{\alpha}$ and $\vec{\beta}$ are optimized, and
    $\Upsilon_{\mathrm{0PN}}$ evaluated using the values
    $\vec{\alpha}$ and $\vec{\beta}$ given by the simplest Newtonian
    (0PN) approximation---the same c.m. correction used in the current
    SXS catalog.  The vertical axis shows the relative improvement in
    going from the Newtonian correction to the optimized correction.
    The dashed diagonal line represents where the comparisons are
    equal---the ``$x=y$'' line.  Optimization improves the results for
    the great majority of systems by less than 1\%.  The three
    exceptions to this rule are particularly long systems.}
\end{figure}

We can also actively optimize $\Upsilon$ over the values of
$\vec{\alpha}$ and $\vec{\beta}$.  The results are shown in
Fig.~\ref{fig:upsilon_comparison_opt}.  Naturally, $\Upsilon$ improves
in every case because it is specifically being optimized.  In
Fig.~\ref{fig:upsilon_comparison_opt}, we see the pattern that the
vast majority of systems are changing by small fractions.  In this
case, there are just three systems in which $\Upsilon$ changes at the
percent level.  These are some of the same systems that changed the
most in going from the raw data to the 0PN-corrected data.  These
particular systems also happen to be extremely long, with significant
overall accelerations during the inspiral.  This suggests that the
corrections will be sensitive to the precise span of times over which
the corrections are being made, which may explain why they continue to
change so much by optimization. However, as discussed in
Sec.~\ref{sec:titf}, changing the beginning and ending fractions does
not significantly change $\vec{\alpha}$ and
$\vec{\beta}$. Nevertheless, the overall scale of the changes seen in
this plot suggests once again that the naive 0PN c.m. correction is
achieving near-optimal results in the vast majority of cases.

\section{Improving the c.m. correction}
\label{sec:ImprovingCOM}

\subsection{Post-Newtonian c.m. definition}
\label{sec:PN}

\begin{figure}
  \includegraphics[width=\columnwidth]{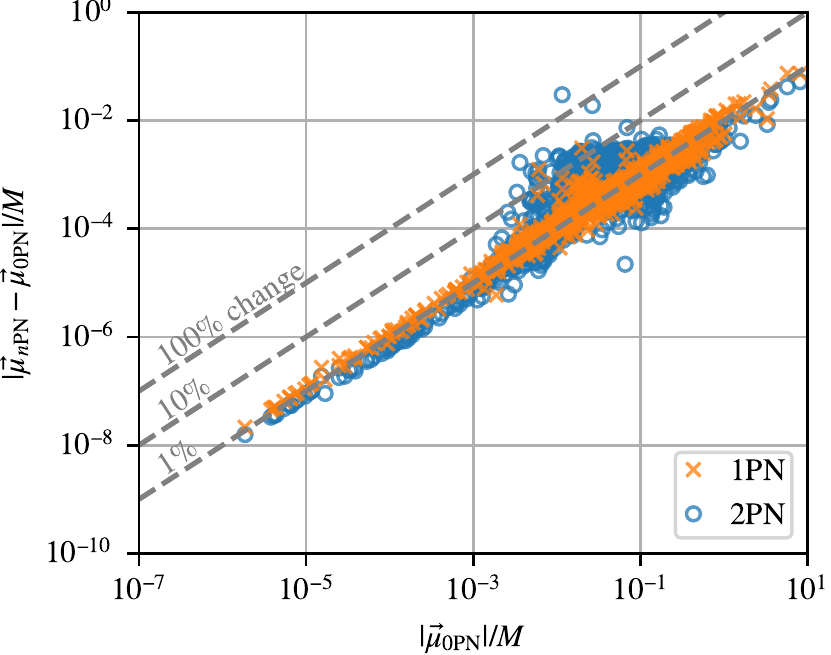}
  \caption{\label{fig:PNcomp}%
    Differences in the maximum displacements between the c.m.
    correction computed using the Newtonian (0PN) c.m. formula and the
    1PN or 2PN c.m. formulas.  The quantity $\vec{\mu}$, defined in
    Eq.~\eqref{eq:mu_definition}, is the largest displacement between
    the origin of coordinates in the simulation and the corrected
    origin.  The post-Newtonian corrections change the c.m. correction
    values by roughly 1\% in the majority of cases.  Systems with
    larger changes are consistent with systems that are sensitive to
    small changes in the end points of integration used to find the c.m.
    correction, as seen in Fig.~\ref{fig:titferror}.}
\end{figure}

\begin{figure}
  \includegraphics[width = \columnwidth]{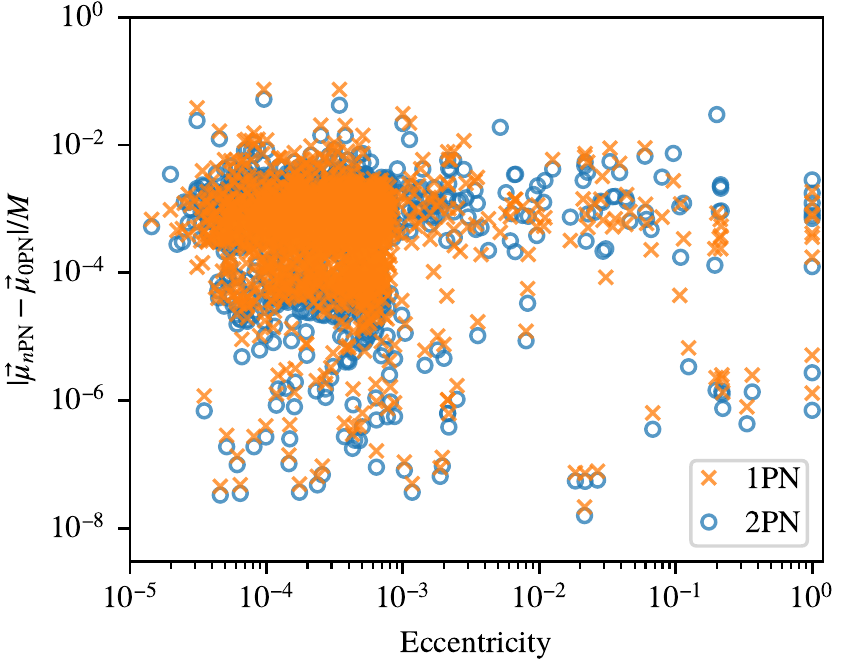}
  \caption{\label{fig:ePN}%
    Differences between the Newtonian c.m. and the 1PN corrected c.m.
    and 2PN corrected c.m. correction values versus the eccentricity of
    the simulation at relaxation time.  No correlations are evident
    between either the 1PN or 2PN correction and eccentricity values.}
\end{figure}

\begin{figure*}
  \includegraphics[width=\columnwidth]{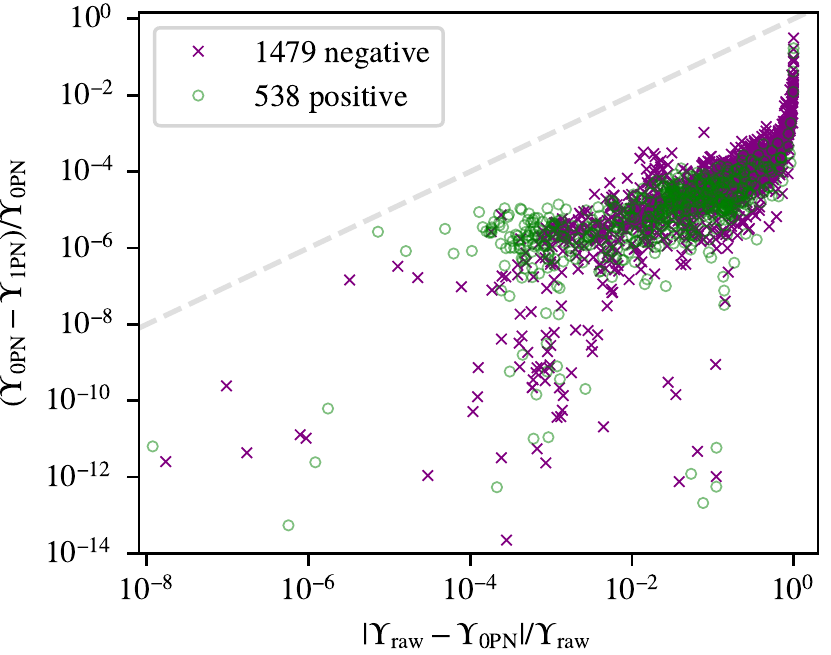}%
  \hfill%
  \includegraphics[width=\columnwidth]{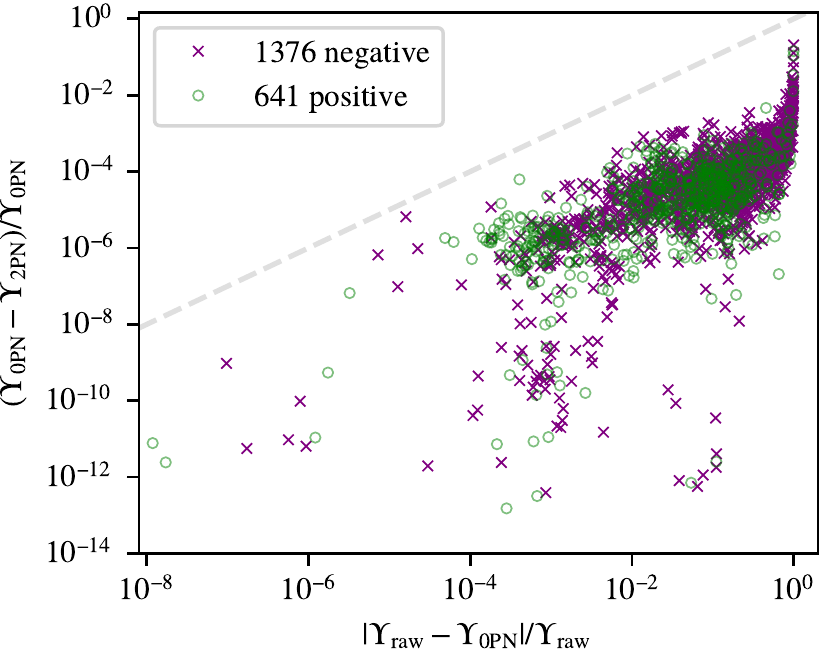}%
  \caption{\label{fig:upsilon_pn_comparisons}%
    Comparison of post-Newtonian contributions for determining the
    center of mass.  These plots show the difference between the value
    of $\Upsilon$ [Eq.~\eqref{eq:objective_function}] resulting from
    the naive 0PN method based on the coordinate trajectories of the
    apparent horizons and the value of $\Upsilon$ when incorporating
    1PN and 2PN effects, as described in the text.  The horizontal
    axes show the relative magnitude of the change when going from the
    raw data to the corrected waveform. The dashed diagonal line in
    both plots represents where the comparisons on the horizontal and
    vertical axes are equal---the ``$x=y$'' line.  In most cases, the
    values actually become significantly \emph{larger} when going from
    the 0PN value to other values.  Those systems are shown as
    crosses, while systems with smaller values are shown as circles.}
\end{figure*}

To characterize the motion seen in the c.m. in the raw simulation data,
an obvious first step to finding a more accurate definition of the c.m.
during the simulation is to try low orders of post-Newtonian (PN)
corrections.  Note that the c.m. should be near the origin of the
coordinate system, with minimal motion around the origin from
linear-momentum recoil, as discussed in Sec.~\ref{sec:linearmom}. 1PN
order corrections are analytically trivial, and are zero for circular
systems. However, SXS simulations are not perfectly circular and so we
investigate 1PN and 2PN order corrections to the c.m.. We implement the
PN corrections given by Eq (4.5) in~\cite{Andrade2001}. This formalism
goes up to 3.5 PN for the c.m. vector in time. The 2PN version of the
expression that we used can be found in Appendix~\ref{sec:PNcor}.

Using this formalism, the effects of the correction on the c.m. are
small but measurable at 1PN and significantly different at 2PN for
many systems.  We are not concerned with the coordinates of the c.m.
itself changing, but of the c.m. correction values changing, as
discussed in Sec.~\ref{sec:COMCorrectionMethod}.  In general, we
assume that the c.m. drifts in a linear fashion and can be corrected
with a translation $\vec{\alpha}$ and boost $\vec{\beta}$, Our results
from the 1PN and 2PN analysis are shown in Fig.~\ref{fig:PNcomp},
which shows the relative difference between $\vec{\alpha}$,
$\vec{\beta}$, and total c.m. displacement
$\vec{\delta}= \vec{\alpha} + \vec{\beta}t_{\text{merger}}$ for the
1PN (top panels) correction to the c.m. and the 2PN (bottom panels)
correction to the c.m.. The 1PN corrections show small changes for most
simulations. However, the 2PN correction shows more sizable changes in
the c.m. correction values, which may indicate that including at least
up to 2PN corrections to the c.m. will give better accuracy either to
the c.m. itself or to the correction factors.

To see any potential correlations with large 1PN or 2PN corrections
and simulation parameters, we compared the relative difference in the
c.m. corrections to the eccentricity $e$ of the system. As shown in
Fig.~\ref{fig:ePN}, no correlations between the magnitude of the 1PN
and 2PN corrections to the c.m. and the eccentricity are apparent,
despite the definition of the 1PN contribution to the c.m. being
dependent on $e$.

Using the method described in Sec.~\ref{sec:Accuracy}, comparisons of
$\Upsilon$, as defined in Eq.~\eqref{eq:objective_function}, between
1PN, 2PN, and the original c.m. correction method (dubbed 0PN), can be
found in Fig.~\ref{fig:upsilon_pn_comparisons}. The striking feature
of these plots is that a significant number of systems actually have
\emph{larger} values of $\Upsilon$ when including either of these
corrections.  While it is reassuring that the majority of systems in
each case only exhibit quite small changes---changes of order
$10^{-4}$ or less in a quantity that already improved significantly
from the raw data---the 1PN and 2PN corrections plots include a large
group of systems that change at the percent level.  These systems also
happen to be the same systems that changed most drastically in going
from the raw data to the 0PN-corrected data (found near the
upper-right corner of the plots), and are particularly biased towards
increasing values of $\Upsilon$.  That is to say, it appears that the
1PN and 2PN corrections do worst for the most extreme systems.  This
should not come as a great surprise, since those systems tend to be the
ones with the most extreme mass ratios and precession, so that
post-Newtonian analysis is also expected to be at its least accurate.

\subsection{Linear-momentum recoil}
\label{sec:linearmom}

Any binary with asymmetric components will emit net linear momentum in
the form of gravitational waves, which will cause a recoil of the
binary itself.  As the system rotates, the direction of recoil will
also rotate, pushing the c.m. roughly in a
circle~\cite{Blanchet:2018yqa, Fitchett1983}.  In principle, this
effect could cause the epicyclical motion apparent in the c.m.
trajectories, which is further characterized in
Sec.~\ref{sec:Epicycles}.  To see if recoil is responsible, we use
methods described in Refs.~\cite{Blanchet:2018yqa, Fitchett1983,
  BoyleEtAl2014, Gerosa2018} to investigate the size of the
linear-momentum recoil implied by the gravitational-wave emission in
these systems.

As shown in the right-column panels of Figure~\ref{fig:ampAndpos}, the
c.m. motion follows an overall linear track with additional epicyclical
motion. The linear motion of the c.m. is well understood, and discussed
in Sec.~\ref{sec:COMCorrectionMethod}. For this analysis, we assume
that the linear part of the c.m. motion may be removed from the data
without loss of information, leaving the epicyclical behavior about
the coordinate origin.

Blanchet and Faye~\cite{Blanchet:2018yqa} consider the motion of the
c.m. from linear momentum flux and the flux of the c.m. itself up to
3.5PN order to calculate the instantaneous c.m. motion induced by these
effects, finding the position of the c.m. relative to its average
location over an orbit to be
\begin{equation}
  \label{eq:commotion}
  \vec{G} = -\frac{48}{5}\frac{G^4}{c^7r^4\omega}
  m_{a}^{2}\, m_{b}^{2}\, (m_{a}-m_{b})\, \hat{\lambda},
\end{equation}
[cf. Eq.~(6.9) in Ref.~\cite{Blanchet:2018yqa}] leading to a circular
motion of the c.m. with radius
\begin{equation}
  \label{eq:p_recoil}
  r_{\text{recoil}} = |\vec{G}| = \frac{48}{5}\frac{G^4}{c^7r^4\omega}
  m_{a}^{2}\, m_{b}^{2}\, (m_{a}-m_{b})
\end{equation}
for a system comprised of nonspinning BHs in a circular orbit. Here,
$r = |\vec{x}_a-\vec{x}_b|$ is the distance between the two black
holes, $\omega = \sqrt{GM/r^3}$ is the Newtonian orbital frequency,
and $\hat{\lambda}$ is the unit vector in the direction of motion of
$m_{a}$ [cf. Eq.~\eqref{eq:unit_vector}].  An earlier analysis using a
simplified model and lower-order approximations can be found in
Ref.~\cite{Fitchett1983}.

\begin{figure*}
  \includegraphics[width=\textwidth]{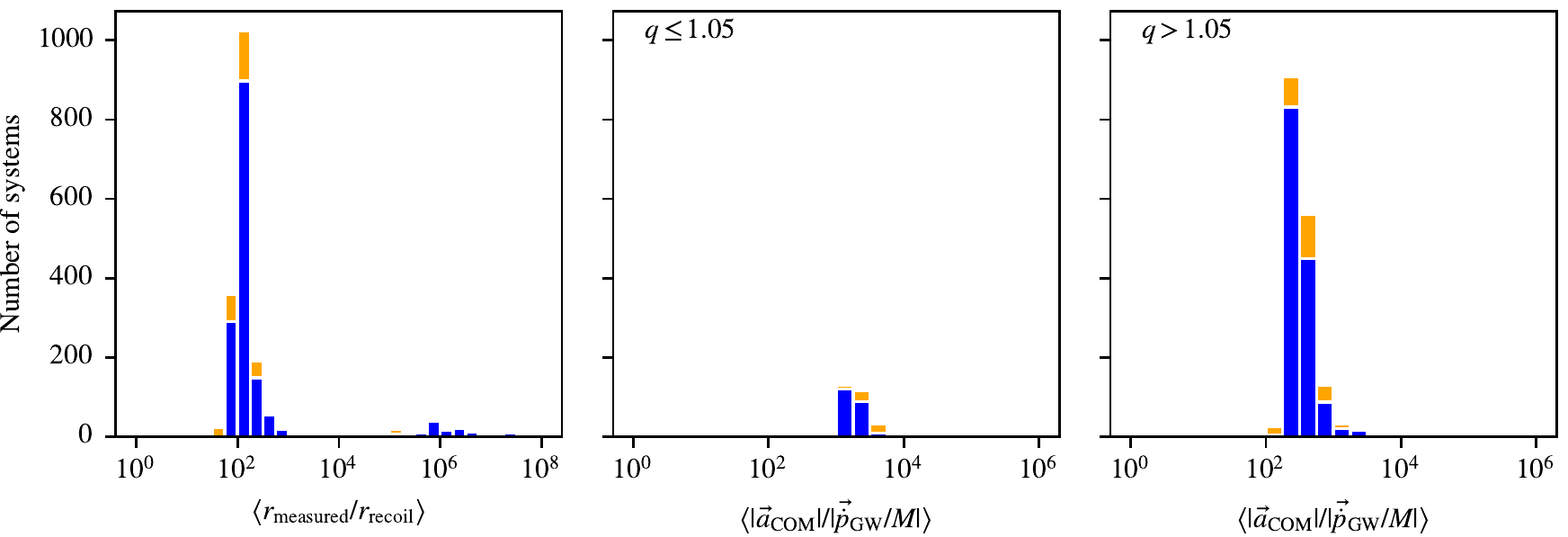}
  \caption{\label{fig:c.m.radius}%
    Comparing measured c.m. motion to motion caused by emission of
    linear momentum carried away by gravitational waves.  The panel on
    the left shows the average ratio of the measured radius of c.m.
    motion, given by the time-averaged magnitude of the c.m. epicycles
    and stated in Eq.~\eqref{eq:rmeasured}, to the radius expected
    from leading-order calculations given by Eq.~\eqref{eq:p_recoil}.
    The center and right panels show the average ratio of the measured
    c.m. acceleration, given by the second time derivative of the c.m.
    coordinate positions and stated in Eq.~\eqref{eq:acom}, to the
    acceleration due to asymmetric momentum flux carried by the
    measured gravitational waves---for near-equal mass ratios and
    larger mass ratios, respectively.  Blue indicates runs using the
    initial-data method described in Ref.~\cite{Ossokine2015}, orange
    indicates runs using the previous initial-data method.  In every
    case, the measured motion is at least an order of magnitude larger
    than the motion expected from gravitational-wave recoil.}
\end{figure*}

Subtracting the motion described by Eq.~\eqref{eq:p_recoil} out of the
c.m. motion and comparing the radius of the measured motion to that of
Eq.~\eqref{eq:p_recoil} immediately shows that the measured c.m.
motion---specifically the epicyclical motion---is significantly larger
than what can be explained by linear-momentum and c.m. reactions for
all simulations.  Figure~\ref{fig:c.m.radius} shows a comparison across
the SXS public waveform catalog for the ratio between the measured c.m.
radius about the coordinate origin and the estimated c.m. radius given
by Eq.~\eqref{eq:p_recoil}. The measured c.m. radius is calculated by
averaging the distance of the c.m. at time $t$ away from the line
$\vec{\alpha}+\vec{\beta}t$ between times $t_i$ and $t_f$:
\begin{equation}
  \label{eq:rmeasured}
  r_{\mathrm{measured}} = \frac{1}{t_f-t_i}\int^{t_f}_{t_i}|\vec{x}_{\text{c.m.}} - (\vec{\alpha}+\vec{\beta}t)|dt.
\end{equation}
The results show that the measured radius is typically hundreds of
times larger than the radius implied by Blanchet and Faye's analysis.

It is also possible to go beyond the analysis in
Ref.~\cite{Blanchet:2018yqa} by using the \emph{measured}
gravitational waves to compute the linear-momentum flux, and compare
the acceleration that would cause to the measured acceleration of the
c.m. in the simulation.  The acceleration of the c.m.,
$\vec{a}_{\text{c.m.}}$ was calculated by taking two time derivatives
of the coordinate position of the c.m. after removing the linear
motion:
\begin{equation}
  \label{eq:acom}
  \vec{a}_{\text{c.m.}} = \frac{d^2}{dt^2} \left(\vec{x}_{\text{c.m.}} - (\vec{\alpha}+\vec{\beta}t)\right).
\end{equation} 
The linear-momentum flux may be calculated from the gravitational
radiation as~\cite{Ruiz2007}
\begin{equation}
  \label{eq:MBlinearmom}
  \frac{d\vec{p}}{d\Omega\, dt} = \frac{c^2}{G} \frac{R^2}{16\pi}
  \left\vert\frac{dh}{dt}\right\vert^2 \hat{r},
\end{equation} 
where $\hat{r}$ is the direction from the source to the point in
question, $R$ is the distance from the source to the observation
sphere, and $\Omega$ represents all angles on the sphere.  Integrating
over all angles to find the total linear momentum flux
$d\vec{p}/dt = \dot{\vec{p}}$, we need to expand $h$ into SWSHs as
done in Eq.~\eqref{eq:h_YlmSum}. We can also decompose $\hat{r}$
accordingly as
\begin{equation}
  \label{eq:rhat}
  \hat{r} = \sqrt{\frac{2\pi}{3}} (Y_{1,-1} - Y_{1,1},\, i Y_{1,-1}+i Y_{1,1},\, \sqrt{2}Y_{1,0}).
\end{equation}
Then integrating over all angles gives
\begin{multline}
  \label{eq:Ylmlinearmom}
  \dot{p}_j = \frac{c^2}{G} \frac{R^2}{16\pi}
  \sum_{l,l',m,m'} \hat{r}_j^{1,m'-m} \dot{h}^{l,m} \bar{\dot{h}}^{l',m'}
  \sqrt{\frac{3(2l+1)(2l'+1)}{4\pi}}
  \\ \times
  (-1)^{m'}
  \begin{pmatrix} l&l'&1\\m&-m'&m'-m \end{pmatrix}
  \begin{pmatrix} l&l'&1\\2&-2&0\end{pmatrix},
\end{multline}
where the last two factors are Wigner 3-$\!j$ symbols.  The sum over
$m'$ and most terms in the sum over $l'$ can be eliminated using
properties of the 3-$\!j$ symbols.  Explicit expressions for the
calculation of the linear momentum flux are given in
Appendix~\ref{sec:pcomponents}.  Our calculation of the linear
momentum flux can be found in the open-source package
\texttt{spherical\_functions}~\cite{SphericalFunctions}.

Of course, there are multiple methods for calculating the linear
momentum for a BBH system.  We compared our method with the method
proposed in Ref.~\cite{Gerosa2018} and originally stated in
Ref.~\cite{Ruiz2007}, and found that the two derivations give the same
results to within numerical accuracy.  This is unsurprising given that
the methods are identical up to the choice of notation. In particular,
the core definition of the total linear momentum flux given in
Eq.~\eqref{eq:MBlinearmom} is the same as that given in Eq.(2.11) of
Ref.~\cite{Ruiz2007}.

We compared the linear-momentum flux divided by the total mass of the
system with the measured acceleration of the c.m. and find that the
values for most of the SXS public waveform catalog do not agree.  We
further found that the c.m. acceleration for nonequal mass systems is
consistently larger than the acceleration found through the linear
momentum flux, confirming that most of the c.m. motion is not due to
linear-momentum recoil.

An overview of the ratios between $\vec{a}_{\text{c.m.}}$ and
$\dot{\vec{p}}/M$ can be seen in Figure~\ref{fig:c.m.radius}.  Note that
linear-momentum recoil in equal mass or near-equal mass BBH systems is
expected to be very small, and so these systems are isolated as a
separate case in the middle panel of Figure~\ref{fig:c.m.radius}.

\subsection{Causes of unphysical c.m. motion}
\label{sec:Causes}

If the c.m. motion seen in the SXS BBH simulations cannot be explained
by physical processes such as inclusion of PN terms or linear-momentum
recoil, then why is it there? There are two potential causes for the
appearance of unphysical or erroneous c.m. motion: (i) the presence of
uncontrolled residual linear momentum in the initial data, and (ii)
the emission of unresolved junk radiation at the beginning of the BBH
simulation causing effectively random and unphysical coordinate kicks
to the system. The presence of uncontrolled residual linear momentum
was addressed and partially rectified in Ref.~\cite{Ossokine2015},
leading to a new method of creating initial data for the BBH
simulations.  However, this method was not used for all systems in the
SXS catalog, and does not completely resolve the issues of spurious
translations and boosts even when it is applied.

Another factor that seems to cause c.m. motion starting from early in
the simulations appears to be junk radiation, which is an inherent
part of BBH simulations. It is the radiation emitted when a BBH
relaxes from its initial-data ``snapshot'', which is only an
approximation to the true state of a long binary inspiral at the time
the simulation starts. Junk radiation is physical, in the sense that
if the entire spacetime were actually in the configuration given by
the initial data, it would indeed emit this radiation as the system
evolved. However, it is not \emph{astrophysical}, in the sense that no
real system in the universe is expected to contain this type of
radiation.  Junk radiation contains high frequencies that are largely
unresolved in BBH simulations because of limits of computational power
and time. As the resolution increases in the simulations, more of the
junk radiation is accurately treated. This is a potential cause for
the difference in initial kicks of the c.m., as seen in Fig 2.  Even in
our higher-resolution simulations, not all of the junk radiation is
accounted for. This failure to resolve all of the junk radiation
possibly contributes to the c.m. kicks observed in SpEC BBH
simulations.

Fortunately, kicks from the emission of unresolved junk radiation can
be corrected using the gauge transformations discussed in the previous
section. The large epicyclical motion in the c.m. cannot be accounted
for using a BMS gauge transformation like the linear motion of the
c.m.. The cause for such large, seemingly unphysical epicyclic motion
is unknown, and is left for future work.

\subsection{Epicycle quantification}
\label{sec:Epicycles}
As seen in the right column panels of Figure~\ref{fig:ampAndpos}, the
c.m. motion has both a linear and epicyclic component. The linear
component of the c.m. motion has been discussed and is already
considered in the current c.m. correction technique.

The size of the epicyclic motion in the c.m. cannot be solely explained
by linear-momentum recoil, as discussed in
Sec.~\ref{sec:linearmom}. The leftmost plot in
Figure~\ref{fig:c.m.radius} illustrates that the expected radius of the
epicycles from linear-momentum recoil, even on an approximate basis,
is orders of magnitude smaller than what is actually seen, given by
$r_{\mathrm{measured}}$ defined in Eq.~\eqref{eq:rmeasured}. The
actual size of $r_{\mathrm{measured}}$ is fairly consistent across the
SXS simulation catalog regardless of simulation parameters or initial
data construction, and tends to be between 0.01 and 0.1 with an
average value of 0.026 in simulation units.  There is not a
significant change in the distribution or magnitude of
$r_{\mathrm{measured}}$ between the 0PN, 1PN, or 2PN representations
of the c.m..

We assume that the epicycle motion seen in the c.m. after the
translation and boost are applied is from the calculated c.m.,
$\vec{x}_{\text{c.m.}}$, being displaced from the optimal c.m. by a
small amount. ``Displaced'' here means that we assume
$\vec{x}_{\text{c.m.}}$ is displaced from the optimal c.m. along the
separation vector $\vec{r}_{ab}=\vec{x}_a-\vec{x}_b$, and not in any
other direction.

Regardless of the origin of these epicycles, removing them to
calculate the linear c.m. correction should improve the quality of the
waveforms. Their removal should reduce the error associated with the
averaging done to calculate the translation and boost correction
values.  As discussed regarding the choice of beginning and ending
times in Sec.~\ref{sec:titf}, the presence of large epicycles has the
potential to affect the reproducibility and reliability of the
calculation of $\vec{\alpha}$ and $\vec{\beta}$. Removing the
epicycles, or at the very least minimizing them, to calculate a more
accurate c.m. correction should further reduce mode mixing. Of course,
epicycle motion cannot be completely subtracted from the waveform
itself as this would require an acceleration correction, which is not
an allowed BMS transformation.

To accurately describe the epicycles, we need to define the
corotating coordinate frame. For our simulations, we have three unit
vectors that describe the rotating coordinate frame:
\begin{align}
  \hat{n} &= \frac{\vec{x_a} - \vec{x_b}}{|\vec{x_a} - \vec{x_b}|} = 
            \frac{\vec{r}_{ab}}{|\vec{r}_{ab}|}, \nonumber
  \\
  \label{eq:unit_vector}
  \hat{k} &= \frac{\vec{r}_{ab} \times \vec{\dot{r}}_{ab}}{|\vec{r}_{ab}\times
            \vec{\dot{r}}_{ab}|} = \frac{\vec{\omega}}{|\vec{\omega}|},
  \\
  \hat{\lambda} &= -\hat{n} \times \hat{k},\nonumber
\end{align}
where $\hat{n}$ points along the separation vector $\vec{r}_{ab}$,
$\hat{k}$ points along the orbital angular velocity $\vec{\omega}$,
and $\hat{\lambda}$ points along the direction of rotation.

This leads us to a potential method for epicycle removal. First, we
calculate the estimated spatial translation $\vec{\alpha}$ and boost
$\vec{\beta}$ from the original c.m. $\vec x_{\rm c.m.}$ using the
current averaging method
\begin{equation}
  \label{eqn:epiremove1}
  \vec{c}_{1} = \vec{x}_{\text{c.m.}} - (\vec{\alpha} + \vec{\beta}t).
\end{equation} 
Note that $\langle |\vec{c}_1| \rangle = r_{\rm measured}$ as defined
in Eq.~\eqref{eq:rmeasured}, using the angle-bracket notation to
denote averaging over time, as for the moments of the c.m. position in
Eq.~\eqref{eqn:com_moments}.  We then calculate the corresponding
corotating coordinate frame unit vectors given in
Eq.~\eqref{eq:unit_vector}.  These unit vectors are then applied to
the original c.m. $\vec{x}_{\text{c.m.}}$ as
\begin{equation}
  \label{eqn:epiremove2}
  \vec{c}_{2} = \vec{x}_{\text{c.m.}} - \Delta_n \hat{n} - \Delta_{\lambda} \hat{\lambda} -
  \Delta_k \hat{k},
\end{equation}
where $\Delta_n = \langle \vec{c}_1(t)\cdot \hat{n}(t) \rangle$,
$\Delta_{\lambda} = \langle \vec{c}_1(t)\cdot \hat{\lambda}(t)
\rangle$, $\Delta_k = \langle \vec{c}_1(t)\cdot \hat{k}(t) \rangle$,
are the time averaged projections of the linearly corrected c.m. onto
the rotational coordinate system unit vectors. The epicycle-corrected
c.m. $\vec{c}_{2}$ is then fed back into the averaging method to get
the final spatial-translation $\vec{\alpha}_{\mathrm{epi}}$ and boost
$\vec{\beta}_{\mathrm{epi}}$ values, which can then be applied to the
waveform data.

\begin{figure}
  \includegraphics[width=\columnwidth]{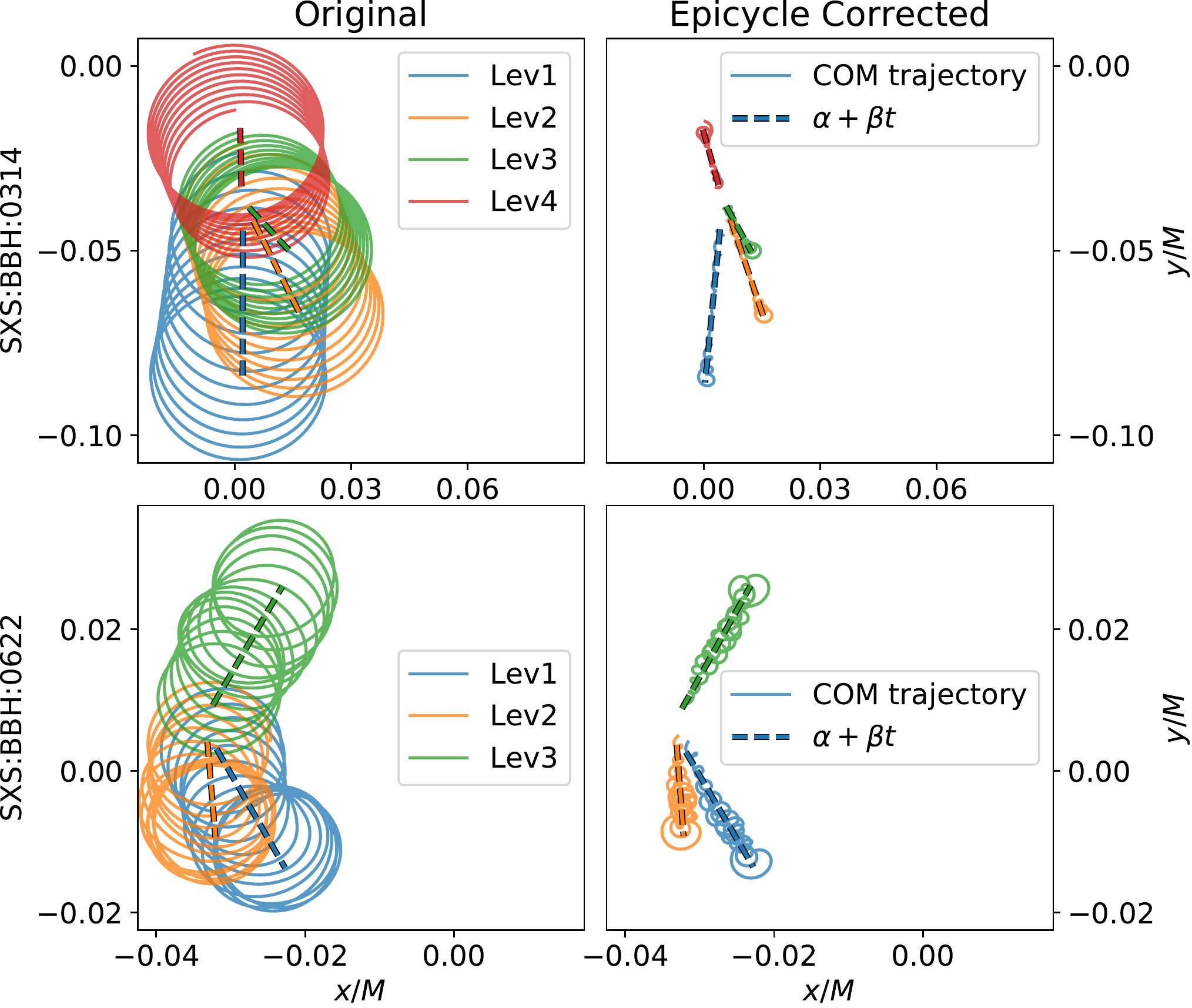}
  \caption{\label{fig:epi_visual}%
    Illustration of epicycle correction for the two simulations
    already shown in Figure~\ref{fig:ampAndpos}.  The left panels show
    the Newtonian center of mass $\vec x_{\rm c.m.}$, whereas the right
    panels show the epicycle corrected $\vec c_2$.  The thick dashed
    lines indicate linear fits to the respective center of mass
    trajectories: $\vec\alpha+\vec\beta t$ (left panels) and
    $\vec\alpha_{\rm epi}+\vec\beta_{\rm epi}t$ (right panels).
    Several numerical resolutions are shown (labeled Lev1 to Lev4),
    and data is plotted only for the time-interval $[t_i, t_f]$, which
    is used for the linear center of motion fits.  }
\end{figure}

A visual representation of our epicycle-correction method is shown in
Figure~\ref{fig:epi_visual}, which uses the spin aligned system
SXS:BBH:0314 and the precessing system SXS:BBH:0622 as sample
cases. As seen in the upper two panels on the right, the removal of
the epicyclic motion as calculated by the time-averaged values
$\Delta_n$, $\Delta_{\lambda}$, and $\Delta_k$ greatly diminishes the
large size of the epicycles and allows for potentially better
optimization of the c.m.-correction values $\vec{\alpha}$ and
$\vec{\beta}$. In these same panels, it can also be seen that not all
of the epicycle motion is removed by our method, and in particular
there are larger deviations towards the beginning and end of the
simulation data, which are a result of our time-averaged method
capturing most but not all of the epicycle motion. Specifically, the
epicycle radius tends to grow with time in most (81\% of) SXS
simulations.

\begin{figure}
  \includegraphics[width=\columnwidth]{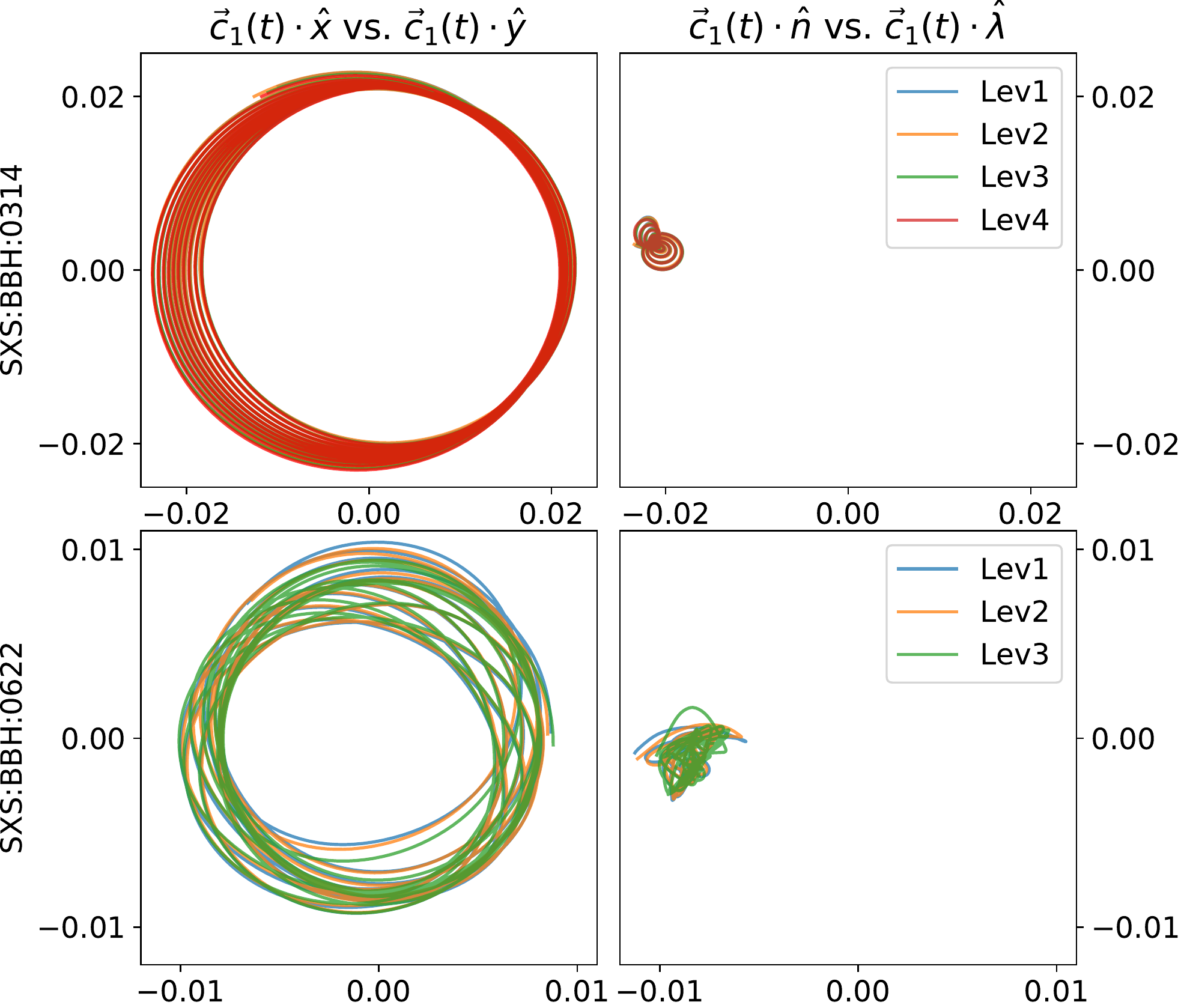}
  \caption{\label{fig:epi_visual2}%
    Contribution to $\vec x_{\rm c.m.}$ which cannot be fitted by a
    linear drift.  {\bf Left panel:} Deviation of Newtonian center of
    mass from a linear motion, i.e. the quantity $\vec c_1$, plotted
    in inertial coordinates. {\bf Right panel:} Projection of
    $\vec c_1(t)$ onto the corotating basis vectors $\hat n$ and
    $\hat\lambda$.  The rotation of $\vec c_1$ around the origin
    visible in the left panels is transformed into a nearly constant
    offset from the origin in corotating coordinates of the right
    panels.  Shown are multiple numerical resolutions (Lev1, Lev2,
    ...) which fall on top of each other, indicating that the
    epicyclic dynamics is numerically resolved and independent of the
    direction of the linear drift.  Data is plotted only for the
    time-interval $[t_i, t_f]$, which is used for the linear center of
    motion fits.  }
\end{figure}

Figure~\ref{fig:epi_visual2} shows $\vec{c}_1$
[Eq.~\eqref{eqn:epiremove1}] and the projections of $\vec{c}_1$ onto
the $\hat{n}$, $\hat{\lambda}$ unit vectors.  These panels show very
similar behavior between resolutions of the same simulation, implying
that the cause of the size of the epicycles is not random, from
initial conditions, or the junk radiation phase, and that the
randomness of the initial kick has been completely removed by the
correction applied in Eq.~\eqref{eqn:epiremove1}.

\begin{figure}
  \includegraphics[width=\columnwidth]{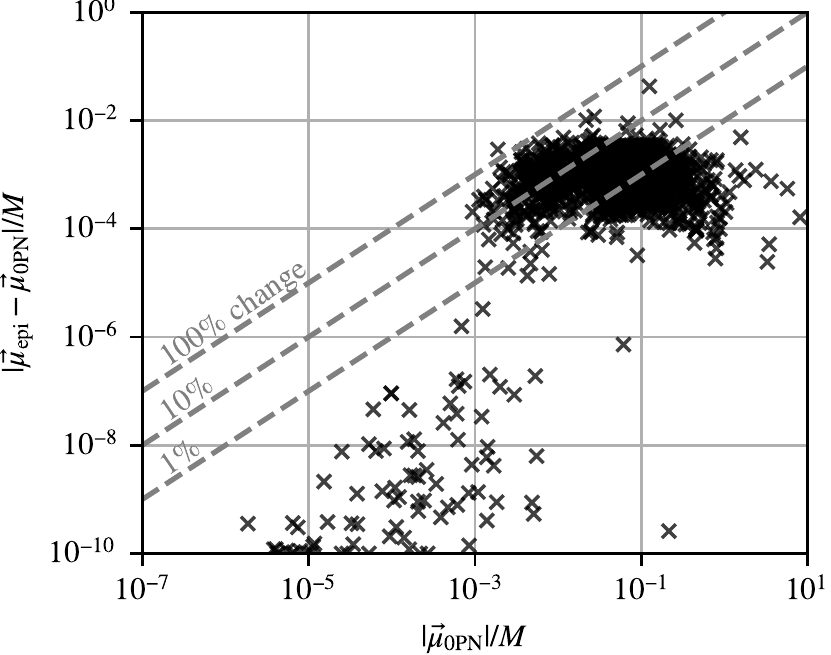}
  \caption{\label{fig:EpicycleRemoval}%
    Change in the size of the c.m. correction, $\vec{\mu}$ as defined
    in Eq.~\eqref{eq:mu_definition}, when removing epicycles before
    fitting for the c.m. correction, as described in
    Eq.~\eqref{eqn:epiremove2}.  These changes are comparable to, but
    almost always smaller than, the changes due to variations in the
    end points of integration as seen in Figure~\ref{fig:titferror}; they
    are also smaller than most of the post-Newtonian corrections seen
    in Figure~\ref{fig:PNcomp}, except for systems changing here by more
    than about 10\%.  }
\end{figure}

Applying this method to the BBH simulations in the SXS public
simulation catalog, we can calculate $\vec{\alpha}_{\mathrm{epi}}$ and
$\vec{\beta}_{\mathrm{epi}}$ c.m. correction values.
Figure~\ref{fig:EpicycleRemoval} compares the usual ``0PN'' c.m.
correction with the epicycle-removed values.  The values plotted
involve $\vec{\mu}$, defined in Eq.~\eqref{eq:mu_definition}, which is
the largest displacement between the origin of coordinates in the
simulation and the corrected origin.  We see that epicycle removal
changes the c.m. correction values at a scale comparable to the changes
caused by varying the end points of integration used to determine the
c.m., as seen in Figure~\ref{fig:titferror}.  The changes due to epicycle
removal are generally somewhat smaller than changes due to varying
end points.  If we remove the epicycles \emph{before} applying those
variations, the changes seen in Figure~\ref{fig:titferror} are reduced
by a typical factor of two---though there is no apparent effect on
roughly 10\% of systems.  Systems that change by more than 10\% in
this figure are also typically changing by more than the
post-Newtonian changes shown in Figure~\ref{fig:PNcomp}.

\begin{figure}
  \includegraphics[width=\columnwidth]{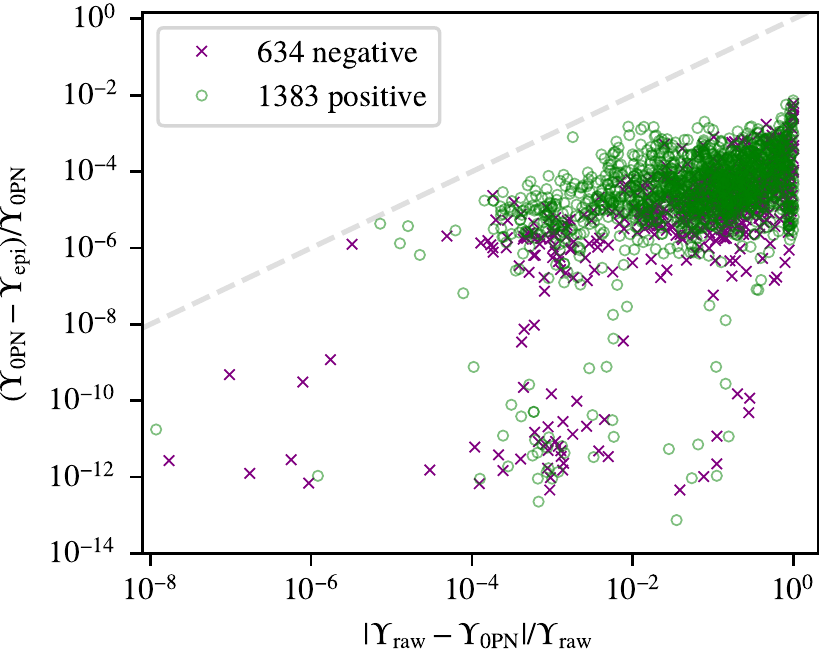}%
  \caption{\label{fig:upsilon_epi_comparison}%
    This plot shows the difference between the value of $\Upsilon$
    [Eq.~\eqref{eq:objective_function}] resulting from the naive 0PN
    method based on the coordinate trajectories of the apparent
    horizons and the value resulting from the epicycle removal method
    described in Sec.~\ref{sec:Epicycles}. The horizontal axis shows
    the relative magnitude of the change when going from the raw data
    to the corrected waveform, while the vertical axis shows the
    change when incorporating the epicycle corrections.  In most
    cases, the values of $\Upsilon$ actually become significantly
    \emph{larger} when going from the 0PN value to the epicycle
    corrected value.  Those systems are shown as crosses, while
    systems with smaller values are shown as circles.}
\end{figure}

Using the method outlined in Sec.~\ref{sec:Accuracy}, we also find
that we cannot reliably conclude that the epicycle correction actually
makes a significant improvement in the waveforms.  The results of the
$\Upsilon$ comparisons between the original c.m. correction method
outlined in Sec.~\ref{sec:SOMCorrectionMethod} and the epicycle
removal method in this section can be found in
Figure~\ref{fig:upsilon_epi_comparison}. This plot shows that
approximately $30\%$ of simulations improve using the epicycle method,
and approximately $70\%$ get worse. This is not enough of a benefit to
warrant the use of the epicycle removal step in all simulations, and
implies that the epicycle removal, at this stage of BBH simulations,
is an unnecessary step in calculating and applying the c.m. correction,
a somewhat disappointing conclusion.

As mentioned in Sec.~\ref{sec:titf}, epicycles are also a potential
source of instability regarding choice of beginning and ending times
$t_i$, $t_f$.  Initial investigation into how the epicycle removal
method affects changes in the c.m. correction values due to differing
$t_i$, $t_f$ implies that the epicycle removal method does not
diminish changes in the c.m. correction values. This is also an
unintuitive and disappointing result, and may imply that other methods
are required to calculate the c.m. correction values $\vec{\alpha}$ and
$\vec{\beta}$ after epicycle removal or a different method for
epicycle removal entirely. We leave such an investigation to future
work.

\subsection{Position of the c.m.}
\label{sec:compos}

\begin{figure*}
  \includegraphics[width=\textwidth]{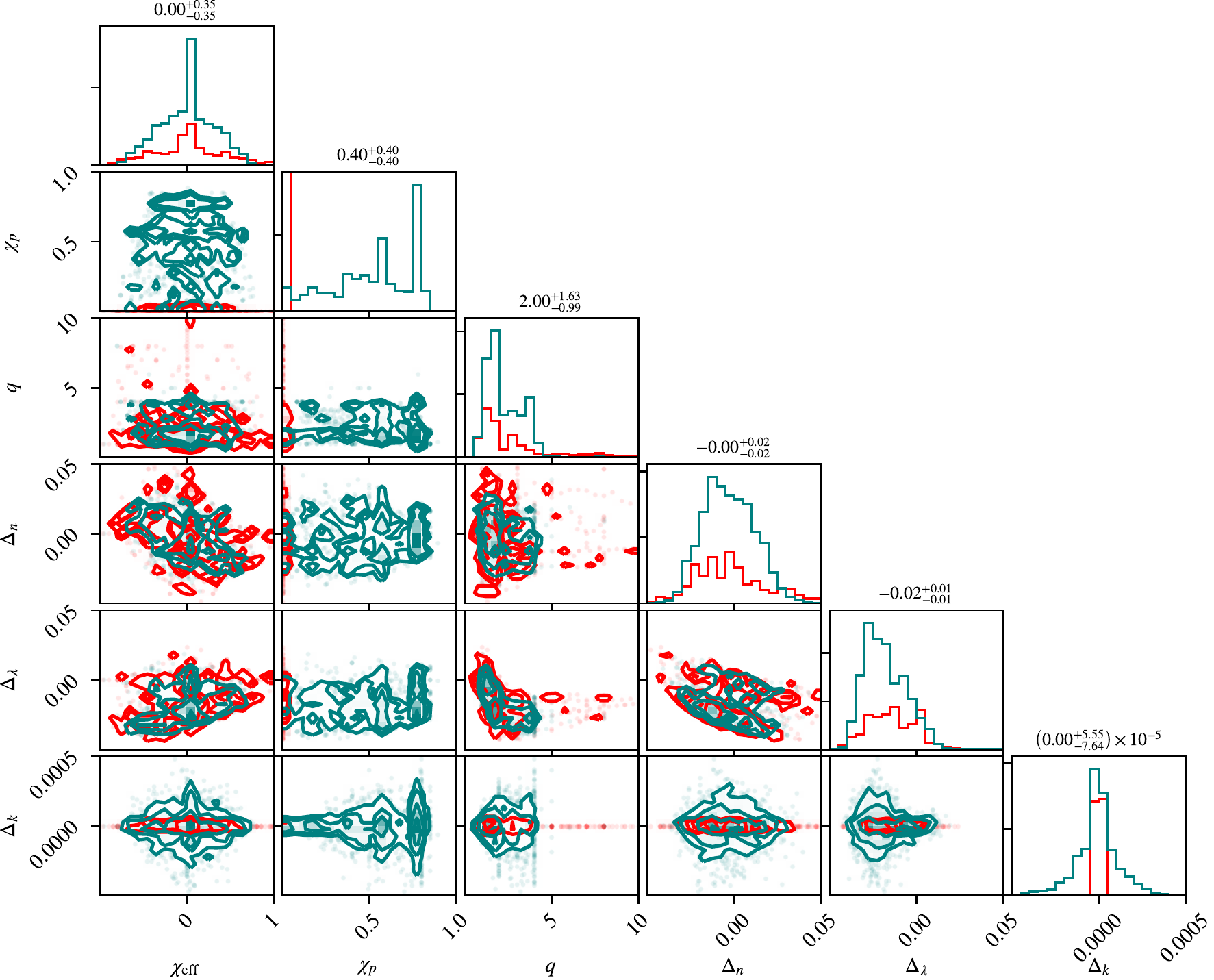}
  \caption{\label{fig:smallcorner} Comparisons of effective spin
    $\chi_{\mathrm{eff}}$ as defined in Eq.~\eqref{eqn:chi_eff},
    effective precessing spin $\chi_p$ as defined in
    Eq.~\eqref{eqn:chi_p}, mass ratio $q$, and the time averaged
    projections of $\vec{c}_1$ onto $\hat{n}$, $\hat{\lambda}$, and
    $\hat{k}$ ; $\Delta_n$, $\Delta_{\lambda}$, $\Delta_k$.  Red
    represents spin aligned simulations, and teal represents
    precessing simulations.  The numbers above each column represent
    the median of each variable over all simulations, with
    superscripts and subscripts giving the offset (relative to the
    median) of the 84th and 16th percentiles, respectively.}
\end{figure*}

During the analysis of the epicycles present in the c.m., we
investigated the position of the c.m. relative to the two black holes.
We still assumed, that $\vec{x}_{\text{c.m.}}$ would lie along the
separation vector between the two black holes or completely in the
rotation direction as given in Eq.~\eqref{eq:commotion}.  As seen in
Figure~\ref{fig:smallcorner}, this is not the case. The c.m. deviates
significantly between the rotation vector and the separation vector
between the two black holes but typically lies in the negative
rotation direction $-\hat{\lambda}$, as predicted by
Eq.~\eqref{eq:commotion} and the analysis in
Ref.~\cite{Blanchet:2018yqa}.  The projection in the
$\pm\hat{\lambda}$ direction, $\Delta_{\lambda}$, averages at
$-0.44r_{\rm measured}$ when considering all simulations in the SXS
Catalog.  On average, the projection of $\vec{c}_1$ into the
$\pm\hat{n}$ direction, $\Delta_n$, is smaller than
$\Delta_{\lambda}$, with an average ratio $\Delta_{\lambda}/\Delta_n$
for spin aligned systems of $-1.43$ and $-2.48$ for precessing
systems. The projection in the $\pm\hat{k}$ direction, $\Delta_k$, is
significantly smaller than $\Delta_{\lambda}$, with an average ratio
$\Delta_{\lambda}/\Delta_k$ of $-3.22\times 10^3$ for spin aligned
systems and $-2.78$ for precessing systems.

Having typically most but not all of the corrected c.m. vector
$\vec{c}_1$ in the direction of $-\hat{\lambda}$ does not have an
obvious cause. This behavior could possibly indicate unaccounted
spin-orbit effects on the c.m., unknown effects from unresolved junk
radiation, or additional gauge effects that cannot be compensated for
with BMS transformations.  Attributes of $\vec{c}_1$ warrant further
investigation, and are left to future work.

Correlations between $\Delta_n$, $\Delta_{\lambda}$, and $\Delta_k$
with pertinent simulation parameters are shown in
Figure~\ref{fig:smallcorner}. This plot shows the correlations between
$\Delta_n$, $\Delta_{\lambda}$, $\Delta_k$, $\chi_{\mathrm{eff}}$,
$\chi_p$, and $q$.  A few notable correlations are apparent, the most
obvious being the correlation between $\Delta_n$ and
$\Delta_{\lambda}$. Most simulations tend to have negative
$\Delta_{\lambda}$ values that grow larger in magnitude with
increasing $\Delta_n$, however there is also a cluster of aligned-spin
simulations with $\Delta_n$, $\Delta_{\lambda}$ values close to
zero. Additionally, $\Delta_{\lambda}$ becomes more negative with
increasing $q$, for $q<5$, and there are some weak correlations
between both c.m. position offsets $\Delta_n$, $\Delta_{\lambda}$ with
$\chi_{\mathrm{eff}}$, but not with $\chi_p$.  $\Delta_k$ does not
appear to have any strong correlations. It is apparent that spin
aligned simulations tend to have $\Delta_k$ values which are much
smaller than for precessing simulations. One possible explanation
for the c.m. to move out of the orbital plane is  momentum flow between
the gravitational fields and black holes~\cite{Keppel09}, however we leave
analysis regarding this mechanism to future work.
It can also be seen in
Figure~\ref{fig:smallcorner} that larger $\Delta_k$ values cluster
around $\Delta_n=0$ and $\Delta_{\lambda} < 0$, which may only be due
to $\Delta_n$ values being symmetric around $0$ and $\Delta_{\lambda}$
values being mostly negative.  No apparent correlations are present
for other simulation parameters.

\section{Conclusions}
\label{sec:Conclusions}

In this work, we have investigated the effects of c.m. motion
waveforms, removed unphysical c.m. motion through allowed gauge
transformations to the waveforms, and have investigated methods for
improving the c.m. correction.  Having unphysical motion in the c.m.
causes mode mixing in the gravitational waveforms, and thus a power
loss from the dominant $(2,\pm 2)$ modes to the less-dominant,
higher-order modes---which is typically visible as amplitude
modulation in the higher-order modes.

We found that the c.m. motion observed in the SXS simulations cannot be
entirely accounted for by PN corrections or linear-momentum recoil.
We also found that the motion of the c.m. does not lie along any one
basis vector describing the rotating coordinate frame as defined in
Eq.~\eqref{eq:unit_vector}, and is offset from the estimated c.m.
within and out of the orbital plane---which is not expected on
physical grounds.

The current method for correcting the c.m. motion uses allowed BMS
transformations, namely a spatial translation and boost that
counteracts the linear motion from the c.m. and removes a large amount
of the mode mixing from the waveform. The translation and boost are
calculated for all simulations at all resolutions, as the c.m. motion
is not consistent between different resolutions of the same system.

We attempted to improve the c.m. correction by developing a method to
remove the large epicycles from the c.m. motion before calculating the
BMS translation and boost. We found that the resulting changes to the
translation and boost values were not significant and did not improve
the waveforms compared to the originally calculated values.

Last, we introduced a complementary method to quantify the effect of
the c.m. correction on the waveforms. We used this method to determine
that PN corrections and the epicycle removal technique did not improve
the c.m. correction transformations, and thus would not further improve
the waveforms or accurately describe the c.m. physically.

Future work includes investigating spin-orbit effects on the c.m. and
the peculiarity of the c.m. position.  Further investigation is
required specifically on the unaccounted for size of the radius of the
epicycles seen in the c.m. motion, which may be due to unknown
spin-orbit or unresolved junk radiation effects, and may be corrected
with additional gauge transformations that minimize the epicycles.

\begin{acknowledgments}
  It is our pleasure to thank Saul Teukolsky for useful conversations
  and guidance on this project.  We also thank Leo Stein for helpful
  insight and suggestions regarding notation. This project was
  supported in part by the Sherman Fairchild Foundation, by NSF Grant
  No. PHY-1606654 at Cornell, and by NSERC of Canada Grant No.
  PGSD3-504366-2017 as well as the Canada Research Chairs Program.

  The computations described in this paper were performed on the
  Wheeler cluster at Caltech, which is supported by the Sherman
  Fairchild Foundation and by Caltech, and on the \texttt{GPC} and
  \texttt{Gravity} clusters at the SciNet HPC Consortium, funded by
  the Canada Foundation for Innovation under the auspices of Compute
  Canada, the Government of Ontario, Ontario Research Fund--Research
  Excellence, and the University of Toronto.
\end{acknowledgments}

\appendix 
\section{Spin-Weighted Spherical Harmonics}
\label{sec:SWSH}

Spin-weighted spherical harmonics (SWSHs) are typically used to
generalize the well-known standard spherical harmonics. Specifically,
SWSHs provide a decomposition of general spin-weighted spherical
functions (SWSFs) into a sum of SWSHs.  Spin-weighted spherical 
functions themselves provide a
vital way to study waves radiating from bounded regions, and so have
an obvious and important application in gravitational-wave astronomy,
which is the focus of this work. Spin-weighed spherical functions 
play two key roles in this
field: (1) describing the magnitude of the wave given any direction of
emission or observation, and (2) providing polarization information.
There are a number of subtleties in defining SWSFs and hence SWSHs,
including dependencies on the chosen coordinate system and the
explicit definition of SWSFs.

The spin weight of a function is defined by how it transforms under
rotation of the spacelike vectors $\Re (m^\mu)$, $\Im(m^\mu)$ where
$m^\mu$ is a complex null vector tangent to the coordinate sphere
$S^2$. The rotation of these spacelike vectors is given by
\begin{equation}
  (m^\mu)' = e^{i \Psi} m^\mu .
\end{equation}
A function $\eta$ is then said to have a spin weight $s$ if it
transforms as
\begin{equation}
  \eta' = e^{s i \Psi}\eta.
\end{equation}
In the case of gravitational waves, the metric perturbation $h$ has a
spin weight of $-2$ \cite{Newman1966,Brown2007} and this decomposition
has been used in numerical relativity extensively.

The classic definition of SWSHs~\cite{Newman1966} writes the functions
in terms of spherical coordinates for $S^2$, giving them as explicit
formulas using polar and azimuthal angles $(\theta, \phi)$ and using
two integer variables that define the order of spherical harmonic to
be used. The convention used here and in SXS is $Y_{l,m}$.

Spin-weighted spherical harmonics are thus classically defined as
\begin{equation}
  \label{eq:swshdef}
  _sY_{l,m} = \begin{cases}
    \left[ \frac{(l-s)!}{(l+s)!}\right]^{1/2}\eth^s Y_{l,m}, & 0\leq s\leq l,\\
    (-1)^s \left[\frac{(l+s)!}{(l-s)!}\right]^{1/2} \bar{\eth}^{-s}
    Y_{l,m}, & -l\leq s\leq 0,
  \end{cases}
\end{equation}
where $\eth$ is effectively a covariant differentiation operator in
the surface of the sphere.  $\eth$ is defined \cite{Newman1966} as
\begin{equation}
  \eth \eta = -(\sin\theta)^s \left\{ \frac{\partial}{\partial \theta}
    + \frac{i}{\sin \theta} \frac{\partial}{\partial \phi}\right\} \left\{ (\sin \theta)^s \eta \right\}
\end{equation} 
when operating on some function $\eta$ that has a spin weight $s$.

The above classic method inherits an unfortunate dependency on the
chosen coordinates.  In particular, SWSHs \emph{cannot} be written as
functions on the sphere $S^{2}$; at best they can only be written as
functions on \emph{coordinates of $S^{2}$}.  As such, SWSHs as defined
in Eq.~\eqref{eq:swshdef} do not transform among themselves under
rotation of the sphere (or, equivalently, rotation of coordinates of
the sphere).  That is, a SWSH in a given coordinate system cannot
generally be expressed as a linear combination of SWSHs in another
coordinate system. A more correct method for defining SWSFs that does
not inherit these coordinate-system dependencies is to represent them
as functions from $\Spin{3} \isomorphic \SU{2}$, which in turn maps
onto $S^2$~\cite{BoyleSWSH}. By forming a representation of
$\Spin{3}$, SWSHs defined in this way do transform among themselves
and still agree with the classic definition.  SWSHs may then be
defined as
\begin{equation}
  \label{eq:newswshdef}
  _sY_{l,m}(\textbf{R}) := (-1)^s \sqrt{\frac{2l+1}{4\pi}} \mathfrak{D}^{(l)}_{m,-s}(\textbf{R}),
\end{equation} 
where $\mathfrak{D}$ is a Wigner matrix, which are representations of
the spin group, and $\textbf{R}$ is the $\Spin{3}$ argument.  Taking
$\textbf{R}$ to be in the unit-quaternion representation of
$\Spin{3}$, the $\mathfrak{D}$ matrices may be expressed as
\begin{align}
  \mathfrak{D}^{(l)}_{m',m}(\textbf{R})
  =& \sqrt{\frac{(l+m)!(l-m)!}{(l+m')!(l-m')!}}
     \sum_{\rho=\rho_{1}}^{\rho_{2}} \binom{l+m'}{\rho} \binom{l-m'}{l-\rho-m} \nonumber\\
   & \quad (-1)^\rho R_s^{l+m'-\rho}\bar{R}^{l-\rho-m}_sR^{\rho-m'+m}_a\bar{R}_a^\rho,
\end{align}
where $\rho_{1} = {\max(0, m'-m)}$, $\rho_{2} = {\min(l+m', l-m)}$,
and $R_s$ and $R_a$ are the geometric projections of the quaternion
into symmetric and antisymmetric parts under reflection along the $z$
axis, which are essentially complex combinations of components of the
quaternion:
\begin{align}
  \label{eq:quaternion_parts}
  R_{s} \defined R_{w} + \i R_{z} \qquad \text{and} \qquad R_{a} \defined R_{y} + \i R_{x}.
\end{align}
The definition given in Eq.~\eqref{eq:newswshdef} is consistent with
the definition of SWSHs typically used within the SXS collaboration
and is the assumed formulation for this work.  For further information
on SWSFs and SWSHs, a comprehensive in-depth discussion of the
history, details, and additional formulations of SWSHs can be found
in~\cite{BoyleSWSH}.

\section{Post-Newtonian Correction to the c.m.}
\label{sec:PNcor}
As discussed in Sec.~\ref{sec:PN}, we used the 1PN and 2PN corrections
to the c.m. as outlined in Ref.~\cite{Andrade2001}. The c.m. up to 2PN
order is given in Eq.~(4.5) of Ref.~\cite{Andrade2001} as
\begin{widetext}
  \begin{align}
    \label{eqn:PNcor}
    G^i = m_ay^i_a
    & + \frac{1}{c^2}\left[y^i_a\left(-\frac{Gm_am_b}{2r_{ab}}+\frac{m_av_a^2}{2}\right)\right] \\
    & + \frac{1}{c^4}\Biggl[v^i_aGm_am_b\left(-\frac{7}{4}(n_{ab}v_a)
      - \frac{7}{4}(n_{ab}v_b)\right)+ y^i_1\Biggl( -\frac{5G^2m_a^2m_b}{4r^2_{ab}}
      + \frac{7G^2m_am^2_b}{4r^2_{ab}}+\frac{3m_av_a^4}{8} \nonumber\\
    &\qquad
      + \frac{Gm_am_b}{r_{ab}}\Biggl(-\frac{1}{8}(n_{ab}v_a)^2
      - \frac{1}{4}(n_{ab}v_a)(n_{ab}v_b)+\frac{1}{8}(n_{ab}v_b)^2
      + \frac{19}{8}v_a^2-\frac{7}{4}(v_av_b)
      - \frac{7}{8}v_b^2 \Biggr)\Biggr)\Biggr]
      +a \iff b \nonumber
  \end{align}
\end{widetext}
where the superscript $i$ designates the vector component being
considered; subscripts $a, b$ designate which object is being
considered; $\vec{y}$ is the position of the body being considered;
$r_{ab} = |\vec{y}_a-\vec{y}_b|$ is the distance between body $a$ and
$b$; $\vec{v}$ is the velocity of the body being considered and
likewise $v$ is the magnitude of the velocity; and
$\vec{n}_{ab}=\vec{r}_{ab}/r_{ab}$. Parentheses here represent the
scalar product of the interior values, e.g.,
$(n_{ab}v_b) = \vec{n}_{ab}\cdot\vec{v}_b$.

Note that this representation of the c.m. position does not include an
overall division by the total mass of the system, and so our
calculations deviate from Eq.~\eqref{eqn:PNcor} only by including an
overall denominator $M=m_a+m_b$.

\section{Linear Momentum Flux from $h^{l,m}$ modes}
\label{sec:pcomponents}

As mentioned in Sec.~\ref{sec:linearmom}, our calculation of the
linear momentum flux from the simulations is based on the formalism
outlined in \cite{BoyleEtAl2014}, which uses the SWSH structure of the
gravitational strain $h$. Starting from Eq.~\eqref{eq:Ylmlinearmom},
which gives the general form of the linear momentum flux in $h^{l,m}$
modes, we can evaluate the components of the linear momentum flux:
\begin{widetext}
  \begin{align}
    \dot{p}_x =& -\frac{R^2}{16\pi} \sum_{l,l',m} (-1)^m \sqrt{\frac{(2l+1)(2l'+1)}{2}} \left(\begin{matrix}l&l'&1\\2&-2&0\end{matrix}\right) \dot{h}^{l,m} \Bigg[\bar{\dot{h}}^{l',m-1}\left(\begin{matrix} l&l'&1\\m&1-m&-1\end{matrix}\right)- \bar{\dot{h}}^{l',m+1}\left(\begin{matrix} l&l'&1\\m&-1-m&1\end{matrix}\right)\Bigg]\nonumber\\
    \dot{p}_y =& -i\frac{R^2}{16\pi} \sum_{l,l',m} (-1)^m \sqrt{\frac{(2l+1)(2l'+1)}{2}} \left(\begin{matrix}l&l'&1\\2&-2&0\end{matrix}\right) \dot{h}^{l,m} \Bigg[\bar{\dot{h}}^{l',m-1} \left(\begin{matrix} l&l'&1\\m&1-m&-1\end{matrix}\right) + \bar{\dot{h}}^{l',m+1} \left(\begin{matrix} l&l'&1\\m&-1-m&1\end{matrix}\right)\Bigg]\nonumber \\
    \dot{p}_z =& \frac{R^2}{16\pi} \sum_{l,l',m} (-1)^{m}\dot{h}^{l,m}\bar{\dot{h}}^{l',m} \sqrt{(2l+1)(2l'+1)}\left(\begin{matrix} l&l'&1\\m&-m&0\end{matrix}\right)\left(\begin{matrix}l&l'&1\\2&-2&0\end{matrix}\right),
  \end{align}
\end{widetext}
where $x,y,z$ refer to the simulation coordinates, with the orbit
typically lying in the $x-y$ plane, $R$ is the distance to the
observation sphere, and $\dot{h}$, $\bar{\dot{h}}$ are the time
derivative and its conjugate of the mode amplitudes. The matrices
throughout the summations are Wigner 3-$\!j$ symbols.

\vfil


\let\c\Originalcdefinition
\let\d\Originalddefinition
\let\i\Originalidefinition

\bibliography{References}


\end{document}